# A Theoretical Study of Process Dependence for Critical Statistics in Standard Serial Models and Standard Parallel Models

Ru Zhang[1], Yanjun Liu[2], and James T. Townsend[2]
[1]University of Colorado, Boulder
[2]Indiana University

## Introduction

An immense number of psychological tasks involve mental operations on various types of perceptual, cognitive, or action entities. Thus, questions concerning whether these operations occur in *parallel* (i.e., simultaneously), in *serial* (i.e., one at a time) or in some more complex fashion arise. Though reaching back to antiquity, the cognitive revolution beginning in the 1950s and 1960s brought renewed interest in such questions (e.g., Murdock, 1971; Sternberg, 1966; Sperling, 1960; Egeth, 1966; Estes & Taylor, 1964; Atkinson, Holmgren, & Juola, 1969).

Before beginning our technical foray, we observe that we often refer either to parallel vs. serial *systems or* parallel vs. serial *models,* which of course, are the mathematical descriptions of the material objects. Our concentration will be on response times (hereafter RTs) since that is the observable variable most frequently brought to bear on this and similar issues.

The third author's formulations, starting in the late 1960s, provided rigorous definitions of serial and parallel systems and related topics, which keyed into their most central aspects (e.g., Townsend, 1969, 1972, 1974). The serial and parallel models employed at the time in the literature tended to be not only much narrower but often defined only verbally. The rough idea of serial systems was that items would be processed in sequence and in an independent and identical-distribution, fashion. That for parallel systems was that they all items began processing simultaneously and that they would be processed independently. Most often, in parallel models, it was also assumed that each item would be processed equally efficiently, no matter how many other processes were going on in parallel. This latter notion pertains to the concept of *capacity* (e.g., Townsend, 1971; Townsend & Ashby, 1978). However, we shall not be focusing on this latter assumption in this investigation, although certain statements may occasionally link up with this concept.

In our more rigorous taxonomy, these special, but important, classes of serial or parallel are now referred to as *standard serial* and *standard parallel models*.

They represent the prototypical types of serial and parallel models and were, and still are, incorrectly in our view, often depicted as the *only* members of their respective classes.

One of the theoretical challenges then and now, was that experimenters often used their data to infer one or other of the general classes of models, when in actuality, they were only testing restricted subclasses. For instance, increasing linear mean RT functions of number of items processed was taken, and still is in some quarters, as implying serial processing and excluding all parallel models. In point of fact, only special serial models make this prediction such as standard serial models. And, there exist intuitive parallel models that readily make this prediction (e.g.,Townsend, 1972, 1974), in particular those suffering from *limited capacity* (e.g., Townsend, 1974; Townsend & Wenger, 2004).

There are now hundreds of experimental and theoretical works on the serial and parallel question (for recent reviews, e.g., Algom, Eidels, Hawkins, Jefferson, & Townsend, 2015; Townsend, Yang, & Burns, 2011; Townsend, Wenger, & Houpt, 2018; Townsend & Wenger, 2004). The rigorous theoretical scaffolding constructed by Townsend and colleagues led not only to understanding of where such issues of parallel vs. serial processing cannot be experimentally adjudicated, but also, more happily, to quite powerful theory-driven methodologies for testing the two types of process arrangements (e.g., Townsend, 1972, 1974; Townsend & Ashby, 1983; Townsend & Nozawa, 1995). Among the most powerful approaches is the topic of this volume, Systems Factorial Technology (SFT; see especially Little, Altieri, Fific' & Yang, 2017). Although the present article does not specifically appeal to SFT, it certainly lies squarely in its tradition, which began back in the 1960s with the aforementioned rigorous taxonomy of process models and scrupulous exploration of parallel and serial systems' properties.

We need to recall the rigorous definitions of serial vs. parallel models. The serial models assume all the processes are executed in an end-to-end manner (Figure 1(a)). If $n$ processes are in the model, there are $n!$ ways to arrange the order of them. The parallel models assume that every process starts simultaneously (Figure 1(b)) but they can terminate at different moments. There are also, like the serial case, $n!$ ways to terminate the $n$ processes in the parallel models. For informal discussion the term *channel,* instead of *process*, will sometimes be associated with parallel systems and models. However, the reader should refer to the more precise mathematical definitions and developments if our meaning is ever unclear.

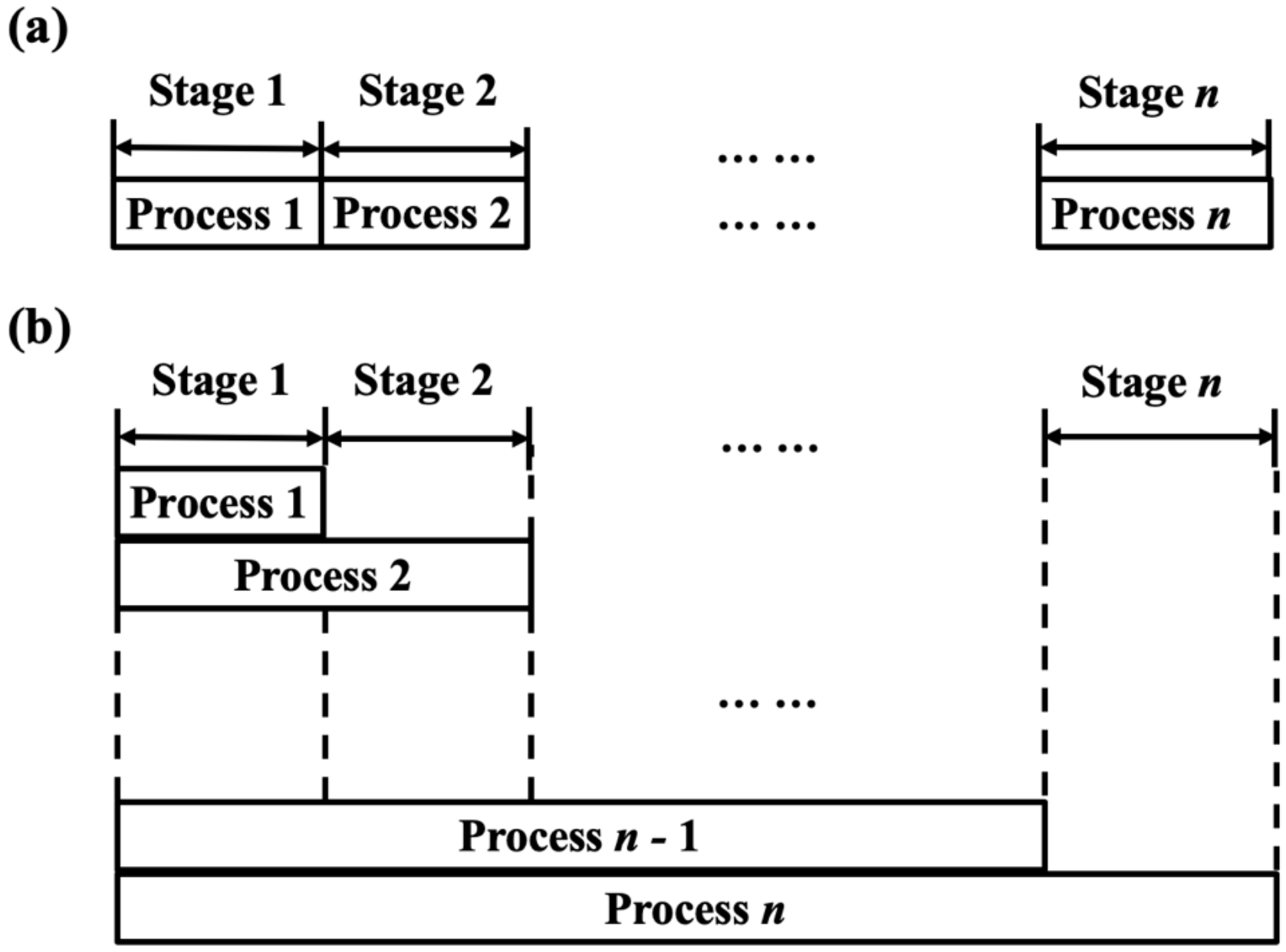

Figure 1. Graphical representations of (a) a serial model and (b) a parallel model.

In this article, we distinguish between the terms *process j* the *j-th process*. The "*j*" in the term "process *j*" represents the name or identity of a process. Whereas "*j*-th process" refers to the position in the processing order. Furthermore, refer to the *processing time* as the actual time to execute a process, which will, of course be distinct for serial and parallel processes

Thus, the variable $z_j$ denotes the true processing time for item going by process *j*, where $j = 1, \ldots, n$. The "*j*" in the term "the *j*-th process" indicates that there are *j* - 1 processes that have finished before that process. In Figure 1, process *j* and the *j*-th process are the same. But please keep in mind, process *j* might be the *i*-th process, where $i \neq j$. Thus, in Figure 1 process 1 might finish as the *i*-th (*i* = 1, 2,…,*n*-1, *n*), whether in a serial or a parallel system.

**Stage.** *Stage j* spans the interval from the end of the (*j* - 1)-th process to the completion of the *j*-th process. Sometimes, the term "stage" is used only in serial models, but we also employ it to act as this kind of descriptive statistic. Nonetheless, for serial models, each process *is* a stage. In Figure 1(a), process 1 is stage 1, process 2 is stage 2,…, and process *n* is stage *n*. For parallel models, as for serial models, a process will typically include several stages. For instance, process 2 in Figure 1(b) includes stage 1 and stage 2 and process *n* includes stage 1, stage 2, to stage *n*.

**Intercompletion time.** The *intercompletion time* $T_j$ (also named the *j*-th intercompletion time) is the time random variable that is spent for stage *j* whether serial or parallel. For the serial models, each processing time is an intercompletion time: In Figure 1(a), $T_1 = z_1$, $T_2 = z_2$,…, $T_n = z_n$. In contrast, the intercompletion time is part of a processing time for the parallel models: In Figure 1(b), $T_1 = z_1$, $T_2 = z_2 - z_1$,…, $T_n = z_n - z_{n-1}$.

**Total completion time for a process.** The *total completion time* $\mathbb{T}_j$ is the time that is consumed from the onset of processing to the moment that process *j* is complete. For the serial model in Figure 1(a), $\mathbb{T}_1 = z_1, \mathbb{T}_2 = z_1 + z_2, \dots, \mathbb{T}_n = z_1 + z_2 + \cdots + z_n$. But, for the parallel model in Figure 1(b), $\mathbb{T}_1 = z_1, \mathbb{T}_2 = z_2, \dots, \mathbb{T}_n = z_n$. To sum up so far, $z_i$ will be a total completion time in a parallel system but an intercompletion time in a serial system.

**Total completion time for a stage.** The total completion time $\mathbb{S}_j$ is the time that is consumed from the onset of the system beginning processing, to the moment that stage *j* is complete. For the serial model in Figure 1(a), $\mathbb{S}_1 = z_1$, $\mathbb{S}_2 = z_1 + z_2$,…, $\mathbb{S}_n = z_1 + z_2 + \cdots + z_n$. For the parallel model in Figure 1(b), $\mathbb{S}_1 = z_1$, $\mathbb{S}_2 = z_2$,…, $\mathbb{S}_n = z_n$. $\mathbb{S}_j$ can also be written as the sum of intercompletion times up to stage *j*. For the serial model in Figure 1(a), $\mathbb{S}_1 = T_1$, $\mathbb{S}_2 = T_1 + T_2$,…, $\mathbb{S}_n = T_1 + T_2 + \cdots + T_n$. For the parallel model in Figure 1(b), $\mathbb{S}_1 = T_1$, $\mathbb{S}_2 = T_1 + T_2$,…, $\mathbb{S}_n = T_1 + T_2 + \cdots + T_n$.

Again, to be sure we're all on the same page, note that the intercompletion time $T_j$ and the total completion time for a stage $\mathbb{S}_j$ are defined with respect to the stage in the processing order, whereas the total completion time for a process $\mathbb{T}_j$ is defined with respect to the identity of a process. And, again in this article, we frequently refer the term *processing time*, which is the actual time to execute a process. To stamp in this concept, for a serial model, processing time *is* equivalent to intercompletion time. For a parallel model, processing time *is* equivalent to total completion time.

Both serial models and parallel models can be represented mathematically. With the foregoing definitions in hand, we shall forthwith employ ICT to designate intercompletion time and TCT to refer to total completion times.

By a natural convention, usually a serial model is associated with ICTs and a parallel model with TCTs, although a serial model could be defined in terms of TCTs and a parallel model could be defined in terms of ICTs (for theoretical purposes, it is often desirable to express both classes in terms of ICTs as in Townsend & Ashby, 1983).

A serial model can be written as the product of the probability of a certain serial order of processing and the joint density function of ICTs conditioned on that order.

$$P(I) f_s(T_1 = t_1, \ldots, T_n = t_n | \, I = (i_1, \ldots, i_n)).$$

$t_1, \ldots, t_n$ are realizations of $T_1, \ldots, T_n$, $(i_1, \ldots, i_n) \in Perm(n)$, where $Perm(n)$ is the set of all permutations of the naturals from 1 to *n*, and $P(I)$ is the probability of a particular permutation $I = (i_1, \ldots, i_n)$. For serial models, permutation *I* means that the model starts with process $i_1$ and is connected to the onset of process $i_2$ after process $i_1$ is complete, and so on. In contrast, parallel models can be most naturally written as the joint density function of TCTs of processes:

$$f_p(\mathbb{T}_1 = \tau_1, \ldots, \mathbb{T}_n = \tau_n; I),$$

where $\tau_1, \ldots, \tau_n$ are realizations of $\mathbb{T}_1, \ldots, \mathbb{T}_n$. For parallel models, permutation *I* means that all the processes start simultaneously but process $i_1$ terminates first, process $i_2$ terminates second, and so on.

Without imposing additional assumptions, serial models and parallel models can perfectly mimic each other in many experimental situations. Then, the experimenter finds it impossible to tell them apart (e.g., Townsend & Ashby, 1983, Chapter 14; Houpt, Townsend & Jefferson, 2017). In order to overcome this problem, several candidate assumptions were raised by scientists.

*Selective influence* (Sternberg, 1969) is the most widely used one. It states that manipulation of each factor only influences the process that is associated with that factor. In serial models, the outcome is that mean response times can be shown to be additive functions of the set of factors. Observe that this statement holds even if the serial model falls outside the class of standard serial models. Over the years, the assumption of selective influence was generalized and refined by Townsend and others (e.g., Townsend, 1984; Townsend & Schweickert, 1989; Schweickert & Townsend, 1989); Dzhafarov, 2003; Townsend & Thomas, 1994).

In fact, SFT (Townsend & Nozawa, 1995) was developed to differentiate parallel models from serial models based on that assumption. One can diagnose the nature of the process arrangements according to the sign of the mean interaction contrast of RTs (e.g., Sternberg, 1969; Schweickert, 1978; Townsend, 1984; Townsend & Schweickert, 1985).

Moreover, much more precise assays were made possible by the extension of contrast functions to survival functions of RTs (Townsend & Nozawa, 1995). SFT has been applied and extended and further explored by a plethora of researchers (e.g., Schweickert, Giorgini, & Dzhafarov, 2000; Dzhafarov, Schweickert, & Sung, 2004; Yang, Fific', & Townsend, 2013; Zhang & Dzhafarov, 2015; Little, Altieri, Fific' & Yang, 2017).

One limitation of SFT is that it requires application in a *complete factorial design*, in which each factor has at least two levels (low salience vs. high salience that result in long processing time vs. short processing time). Not every feature or dimension of interest can fulfill this requirement. For instance, the processing time of red color may be neither faster nor slower than green.

Another candidate assumption is that of *within stage independence*, which states that unfinished parallel processes are independently executed within each stage. Since there is only one process in each stage in a serial model, this assumption is only effective in the class of parallel models. Although within stage independence is an important characteristic to know about, it turns out that within stage dependent models can be mathematically transformed to within stage independent models (e.g., see Rao, 1992, pp. 162-163). Thus, within stage independent parallel models and within stage dependent parallel models cannot be discriminated in the absence of direct observability of the within stage dependencies. The next paragraph is critical to understanding the remainder of our developments and is thereby italicized though it includes concepts already introduced.

*In this article, the axioms concerning processing time independence are absolutely central. For the serial models, this assumption is equivalent to across stage independence since the actual processing times are equivalent to the ICTs. In contrast, for the parallel models, processing time of an item or channel is equivalent to the time spent from the very initiation of processing until an individual channel is finished. As introduced earlier, this is a statistic known generically as the TCT for a process. Independence of processing times in parallel models is tantamount to independence of the TCTs for individual processes. In addition, we assume each processing time is identically distributed. By assuming independently and identically distributed (iid) processing times, serial models and parallel models are termed standard serial models and standard parallel models, respectively (e.g., Algom, Eidels, Hawkins, Jefferson, & Townsend, 2015; Townsend, Houpt, & Wenger, 2018).*

Many experimental paradigms are traditionally used to systematically study the serial-parallel issue. Free recall is one of them, though far from the most popular. In this paradigm, subjects are instructed to recall words that belong to a semantic category from their long-term memory (Bousfield & Sedgewick, 1944; Bousfield, Sedgewick, & Cohen, 1954), for instance, naming as many cities in the United States as they could remember. The words are reported successively. It is found that the time interval between two successive responses, that is the ICT, increases as more responses are generated (Murdock & Okada, 1970; Patterson, Meltzer, & Mandler, 1971; Pollio, Kasschau, & DeNise, 1968; Pollio, Richards, & Lucas, 1969, Lohnas, Polyn, & Kahana, 2011; Polyn, Norman, & Kahana, 2009; Sederberg, et al., 2006, 2010).

Various serial and parallel models have been proposed to interpret the behavior of the data that is observed in this paradigm. Here we introduce several important ones. McGill contributed an influential chapter on stochastic processes in psychology, to the 1963 Volume 1 of the Handbook of Mathematical Psychology (McGill, 1963). He accounted for the general temporal characteristics of Bousfield & Sedgewick's (1944) data based on a serial model. His model assumes that only one item could be sampled from a search set and inspected at any time. All the relevant items are assumed to be chosen with the equal chance at each draw. After each draw, the subject examines if the item is a member of the specified category and if the item is not recalled yet. The amount of time for each draw and subsequent check is assumed to be exponentially distributed with the same rate parameter. Intriguingly, McGill's (1963) serial model is mathematically identical to the standard parallel model with exponential processing times.

Vorberg and Ulrich (1987) subsequently generalized McGill's model according to the assumption of unequal accessibility. By allowing some items more easily to be accessed from memory than the others (Shiffrin, 1970), the generalized model removes some minor discrepancies between the data and McGill's model in predicting the number of generated items by a certain time moment. The stochastic representation of Vorberg and Ulrich's (1987) serial model is then found to be

$$P(I)f_s\big(T_1 = t_1, \dots, T_n = t_n \big| \, I = (i_1, \dots, i_n)\big)$$
$$= \left(\prod_{j=1}^{n} \frac{u_{i_j}}{\sum_{l=j}^{n} u_{i_l}}\right) \left[\prod_{j=1}^{n} \left(\sum_{l=j}^{n} u_{i_l}\right) exp\left(-\sum_{l=j}^{n} u_{i_l}\, t_j\right)\right],$$

where $u_{i_j}$ stands for the rate parameter for process (or item) $i_j$ and *n* is the number of recallable target items within the search set. The model predicts that the rate parameter of the *j*-th ICT equals the sum of rate parameters of processes that have not been executed, that are the *j* + 1-th process to the *n*-th process. Note that when the equal accessibility assumption is imposed ($u = u_{i_1} = \cdots = u_{i_n}$), the ICT distribution does not depend on the recall order any more and the conditional joint density function is reduced to

$$f_s\big(T_1 = t_1, \dots, T_n = t_n \big| I = (i_1, \dots, i_n)\big)$$
$$=\prod_{j=1}^{n}(n-j+1)u exp\big[-(n-j+1)ut_j\big],$$

which is the original McGill (1963) model. Note by observing the above equation, McGill (1963)'s model indicates that the ICTs, or equivalently processing times, are not iid. Vorberg and Ulrich (1987) also derived the stochastic representation of the counterpart parallel model with the assumption of unequal accessibility:

$$f_p\big(\mathbb{T}_1 = \tau_1, \dots, \mathbb{T}_n = \tau_n; I = (i_1, \dots, i_n)\big)$$
$$=\prod_{j=1}^{n} u_{i_j} exp\left(-u_{i_j}\tau_{i_j}\right).$$

Observe that the TCTs are stochastically independent of each other in the counterpart parallel model. By imposing the equal accessibility assumption ($u$ = $u_{i_1}$ = ⋯ = $u_{i_n}$), the above equation is reduced to

$$f_p\big(\mathbb{T}_1 = \tau_1, \dots, \mathbb{T}_n = \tau_n; I = (i_1, \dots, i_n)\big)$$
$$=\prod_{j=1}^{n} u exp\left(-u\tau_{i_j}\right),$$

which is McGill(1963)'s model in the parallel model's representation. Observe it is also the mathematical representation of the standard parallel model with exponential processing times.

Following McGill's model, Rohrer and Wixted (1994) derived a function for the mean ICT changes as more items are generated:

$$\overline{T_J} = \frac{1}{u(n-j)},$$

where $\overline{T_J}$ is the mean of the *j*-th ICT. This equation reflects that an ICT is inversely proportional to $n - j$. Thus, the last ICT of a four-item recall should equal the last ICT of a nine-item recall.

Rohrer and Wixted (1994) conducted experiments by asking the subjects to recall the words studied earlier. By manipulating the size of the study word list or/and the presentation time for each word, they found that the hyperbolic ICT growth reflected by the above equation fitted the data well. As mentioned earlier, McGill's serial model is identical to the standard parallel model with exponential processing times. Thus, Rohrer and Wixted (1994)'s study supports the independent and identical processing time assumption if one considers the recall a parallel process.

In this paper we aim to dig deeper into foundational distinctions between standard serial and standard parallel models. This goal leads us directly to investigate the dependence of TCTs, the actual processing times of parallel systems as well as the dependence of ICTs, the actual processing times of serial systems.

As we have mentioned, a primary axiom of a standard serial model is that the ICTs (i.e., the processing time) are independent. In contrast, for a standard parallel model, the TCTs (i.e., the processing times) are independent. Thus, our mission in this work, is to explore the ICT dependence characteristics of standard parallel models and the TCT dependence characteristics of standard serial models.

To be more specific, we investigate the behavior of (conditional) distributions, of the TCT of processes and the ICT in complete generality, that is, without assuming any particular form for the distributions of processing times. We further examine under what condition the theoretically derived behavior from the standard parallel model is consistent with the empirical findings that the ICT grows as a function of output position in free recall experiments.

In a previous publication (Zhang, Liu, & Townsend, 2018), our treatment was limited to two channels or stages in operation. In the current study, we extend our conclusions to the case of arbitrary *n*. In order to facilitate comprehension of the general case, we provide a brief review of findings for the two-process models in the next section. All the theorems, corollaries, and lemmas in this section were reported in our earlier publication (Zhang, Liu, & Townsend, 2018). The readers can access the mathematical proofs and computational simulations associated with those theorems, corollaries, and lemmas from that chapter.

## Standard Two-Process Models

Suppose there are only two processes $a$ and $b$ in the models to realize standard two-process serial models and the comparable standard two-process parallel models.

Let us denote the processing times of $a$ and $b$ as $z_a$ and $z_b$ (recall that they are iid, whether in a serial or a parallel model) and the density function for each as $f$. The corresponding distribution function is labeled as $F$. The survival function, the hazard function, and the cumulative hazard function are represented respectively as

$$S(x) = 1 - F(x),$$

$$h(x) = \frac{f(x)}{S(x)},$$

$$H(x) = \int_0^x h(x)\,dx = -ln[S(x)],$$

where $x$ denotes a temporal variable. The readers should pay attention to the notation. In this paper $S$ always denotes the survival function.

**Standard two-process serial models.** Since two processes are under consideration, the model can be decomposed into two stages. If process $a$ is executed earlier than process $b$, then process $a$ is stage 1. If process $a$ is executed later than process $b$, then process $b$ is stage 1 (Figure 2).

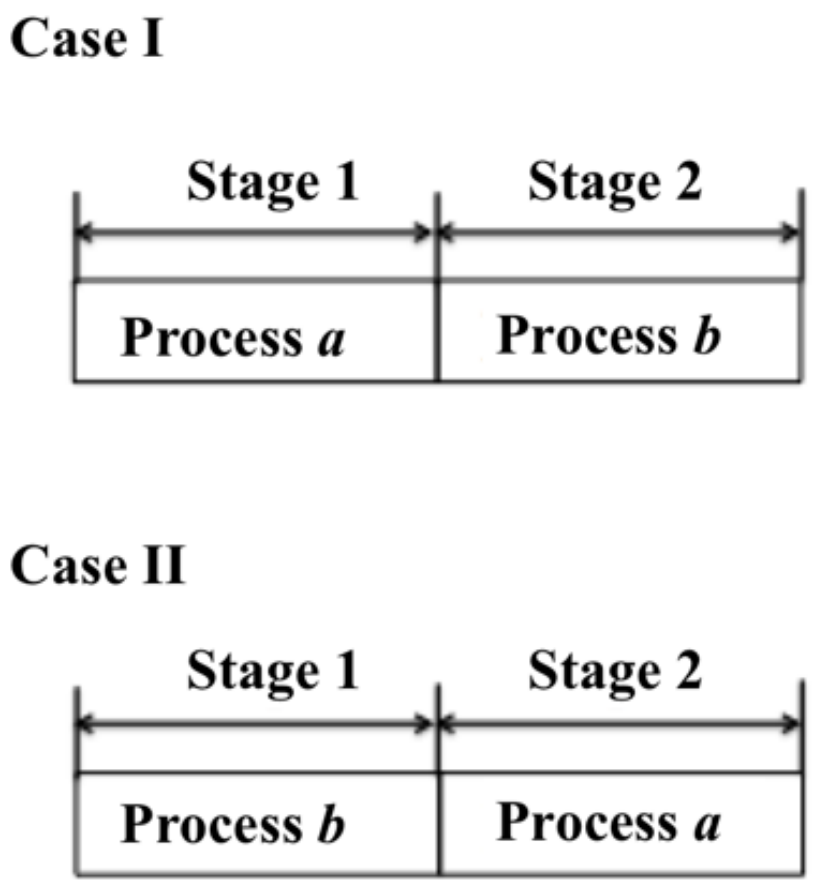

Figure 2. Possible process arrangements of a standard two-process serial model. From "A Theoretical Study of Process Dependence for the Standard Two-Process Serial Models and Standard Two-Process Parallel Models," by R. Zhang, Y. Liu, and J. T. Townsend, in T. Lachmann and T. Weis (Eds.), *Invariances in Human Information Processing* (pp. 117-142), 2018, New York, NY: Taylor & Francis Group. Copyright 2018 by Taylor & Francis Group. Reprinted with permission.

As defined earlier, the ICT $T_1$ is the time that is spent processing in stage 1 and $T_2$ is the time that is spent processing in stage 2. So, for Case I, $T_1 = z_a$, $T_2 = z_b$ and for Case II, $T_1 = z_b$, $T_2 = z_a$. It is apparent that $T_1$ and $T_2$ are iid as $z_a$ and $z_b$ are assumed iid. The TCT $\mathbb{T}_a$ is the time that is consumed from the onset of the model to the moment that process $a$ is completed. The TCT $\mathbb{T}_b$ is the time that is consumed from the onset of the model to the moment that process $b$ is completed. Therefore, for Case I, $\mathbb{T}_a = T_1 = z_a$, $\mathbb{T}_b = T_1 + T_2 = z_a + z_b$ and for Case II, $\mathbb{T}_a = T_1 + T_2 = z_a + z_b$, $\mathbb{T}_b = T_1 = z_b$.

**Standard two-process parallel models.** Since two processes are under consideration (see an example in Figure 3), the TCT for process *a* and the TCT for process *b* are

$$\mathbb{T}_a = z_a,$$
$$\mathbb{T}_b = z_b,$$

respectively. Please note that Figure 3 is an exemplar representation of a standard two-process parallel model in which process $a$ is faster than process $b$. With some non-zero probability, process $a$ will be slower than process $b$ as $z_a$ and $z_b$ are iid. The ICTs in Figures 3 can be represented as

$$T_1 = \mathbb{T}_a = z_a,$$
$$T_2 = \mathbb{T}_b - \mathbb{T}_a = z_b - z_a.$$

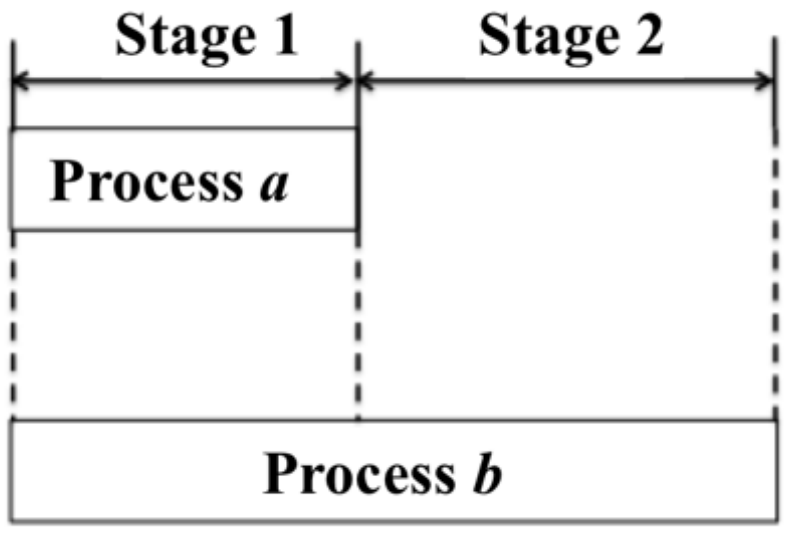

Figure 3. A standard two-process parallel model. From "A Theoretical Study of Process Dependence for the Standard Two-Process Serial Models and Standard Two-Process Parallel Models," by R. Zhang, Y. Liu, and J. T. Townsend, in T. Lachmann and T. Weis (Eds.), *Invariances in Human Information Processing* (pp. 117-142), 2018, New York, NY: Taylor & Francis Group. Copyright 2018 by Taylor & Francis Group. Reprinted with permission.

**Dependence of Total Completion Times, *n* = 2**

As observed earlier, the TCT of a parallel channel (item, etc.) is its processing time and perforce, in a standard parallel model is independent of the processing times of the other channels. An intuition about the potential comparable dependencies in a serial model is the following: condition TCT for process "*b*" on that for "*a*". Now, if "*b*" is done first then, since the TCT for "*a*" is the sum of the two processing times for "*a*" plus that of "*b*", the probability that the TCT for "*b*" is less than $\tau$, given that the sum is already < $\tau$, must be 1. If "*a*" is done first by time $\tau$, the probability that "*b*" also gets done by $\tau$ is greater than its marginal probability. Thus, this qualitative intuition suggests that the TCT in a serial model should be positively dependent. We will learn that this reasoning is faulty in general but correct when only one processing order is allowed.

We move on to perform the actual required computations, comparing the distribution function of $\mathbb{T}_b$ conditional on $\mathbb{T}_a$ versus the marginal distribution function of $\mathbb{T}_b$. That is

$$P(\mathbb{T}_b \leq \tau | \mathbb{T}_a \leq \tau) - P(\mathbb{T}_b \leq \tau). \tag{1}$$

If it is always positive then we conclude that the TCTs in this case are positively dependent in a strong distributional sense, and conversely if the difference is negative. Please note that (1) considers two possible permutations as illustrated in Figure 2.

It was proven by Townsend & Ashby (1983, Page 73-74), if the processing times $z_a$ and $z_b$ (or the ICTs $T_1$ and $T_2$) in a two-process serial model are iid and follow exponential distributions, then $(1) > 0$ for $\tau > 0$. This result is compatible with the above intuition. However, we recently found (1) $> 0$ does not hold for all processing time distributions (Zhang, Liu, & Townsend, 2018).

**Theorem 1.** For a standard two-process serial model, $P(\mathbb{T}_b \leq \tau | \mathbb{T}_a \leq \tau) - P(\mathbb{T}_b \leq \tau)$ can be either positive or nonpositive for $\tau > 0$.

Corollary 2 states if only one permutation is allowed in the investigated system, dependence of TCTs is non-negative, that is $(1) \geq 0$. We also compute the covariance of TCTs. The result is presented in Lemma 3.

Corollary 2 and Lemma 3 correspond to our original intuitions and the Townsend & Ashby (1983) theorem when the distributions are gamma (Erlang). The covariance result in Lemma 3 is especially pleasing.

**Corollary 2.** For a standard two-process serial model, $P(\mathbb{T}_b \leq \tau | \mathbb{T}_a \leq \tau) - P(\mathbb{T}_b \leq \tau)$ is nonnegative for $\tau > 0$ if only the Case I arrangement or only the Case II arrangement is allowed in the investigated system.

**Lemma 3.** For a standard two-process serial model, if only the Case I arrangement or only the Case II arrangement is allowed, $Cov(\mathbb{T}_a, \mathbb{T}_b) = Var(T_1) > 0$.

We know by the very definition of standard parallel models, that the TCTs are independent. However, we include the obvious statements for ease of reference. It is therefore listed that $(1) = 0$ and this statement is presented in as Theorem 4. We also compute the covariance of TCTs for standard two-process parallel models. The result is presented in Lemma 5.

**Theorem 4.** For a standard two-process parallel model, $P(\mathbb{T}_b \leq \tau | \mathbb{T}_a \leq \tau) - P(\mathbb{T}_b \leq \tau) = 0$ for $\tau > 0$.

**Lemma 5.** For a standard two-process parallel model, $Cov(\mathbb{T}_a, \mathbb{T}_b) = 0$.

Standard two-process serial models and standard two-process parallel models can be differentiated according to Theorem 1 and Theorem 4: $P(\mathbb{T}_b \leq \tau | \mathbb{T}_a \leq \tau) - P(\mathbb{T}_b \leq \tau)$ cannot always be zero for a standard two-process serial model; while as for a standard two-process parallel model, the function maintains zero along the axis of $\tau$. One can also differentiate the two models according to Lemma 3 and Lemma 5.

**Dependence of Intercompletion Times, *n* = 2**

The ICTs, and therefore the processing times, in a standard serial model are assumed iid. Therefore, the empirical finding that as the number of stages already completed increases, the ICTs increase, cannot be accounted for by standard serial models. In contrast, standard parallel models can account for this phenomenon as noted by McGill (1963) and Vorberg & Ulrich (1987). The intuition of course, is that the later stages included fewer and fewer parallel processes still to complete and therefore the probability that the minimum time for one of these remaining processes to finish inevitably lengthens. However, it is so far unknown as to whether this behavior is characteristic of all standard parallel models. We shall learn that it is not. First, however, we will investigate a related, but distinct question.

Without loss of generality, we label the process completed earlier process *a* and the other is labeled as process *b* in a standard two-process parallel model. Recall that the processing times $z_a$ and $z_b$, or equivalently $\mathbb{T}_a$ and $\mathbb{T}_b$, are assumed iid. Now let us label the ICTs for stage 1 and stage 2 as $T_a$ and $T_b$, where

$$T_a = \mathbb{T}_a = z_a,$$
$$T_b = \mathbb{T}_b - \mathbb{T}_a = z_b - z_a.$$

One topic of interest is how the ICTs act in later stages as a function of the magnitude of the earlier stages. For *n* = 2, this simply suggests we investigate the likelihood that the second stage ICT $T_b$, is not completed by time *t*, given that the first stage is entirely completed. We thus explore the survival function of the ICT $T_b$ conditional on the processing time of stage 1: $P(T_b > t | z_b > z_a)$, where $t > 0$. Interestingly, it is found that the behavior of $P(T_b > t | z_b > z_a)$ depends on the hazard function of processing time $h$.

**Lemma 6.** For a standard two-process parallel model, if the hazard function $h$ for an arbitrary channel is non-increasing, then $P(T_b > t | z_b > z_a)$ is non-decreasing as $T_a$ is increased.

This result begins to tie the behavior of subsequent ICT times with the hazard function. The finding that they stochastically lengthen as a function of the previous finishing time, if the hazard function is non-increasing, makes intuitive sense.

A separate issue, and one more directly related to the question of whether successive ICTs tend to grow longer, is "How does the later stage duration (stage 2) compare with the earlier stage duration (stage 1) in a standard two-process parallel model?" Let us denote the ratio of the hazard functions:

$$\alpha(t, T_a + t) = \frac{h(T_a + t)}{h(t)}.$$

The survival function at stage 1 in the standard two-process parallel model is the product of two survival functions $S^2(t)$. The survival function at stage 2 is

$$\begin{aligned} &P(T_b > t | z_b > z_a) \\ &= P(T_b + T_a > t + T_a | z_b > z_a) \\ &= P(z_b > t + T_a | z_b > z_a) \\ &= \frac{S(T_a + t)}{S(T_a)}. \end{aligned}$$

We then investigated $S^2(t)$ vs. $\frac{S(T_a+t)}{S(T_a)}$. If the survival function from stage 1 to stage 2 is increasing, that is $S^2(t) - \frac{S(T_a+t)}{S(T_a)} < 0$, this trend is then consistent with the empirical finding that the ICT grows as the number of stages grows, in a strong distributional sense. Theorem 7 provides an exact condition for the survival function to increase from stage 1 to 2. Corollary 8 states that standard two-process parallel models with a concave or linear cumulative hazard function $H(t)$ predict an increasing survival function from stage 1 to stage 2, for arbitrary t.

**Theorem 7.** In a standard two-process parallel model, if $\alpha(t, T_a + t) < 2$, then $S^2(t) - \frac{S(T_a+t)}{S(T_a)} < 0$ so that the survival function of ICT is increasing from the first stage to the second stage; if $\alpha(t, T_a + t) \geq 2$, then $S^2(t) - \frac{S(T_a+t)}{S(T_a)} \geq 0$ so that the survival function of ICT is non-increasing from the first stage to the second stage.

Thus, we have the intriguing and reasonable prediction that any standard parallel model whose hazard function ratio α is less than the number 2 (i.e., $\alpha < 2$) will evidence increasing conditional survivor functions and hence result in the longer ICT in the second stage than the first stage. This result has an elegant consequence that can be expressed in terms of the integrated hazard functions, as visited in Corollary 8.

**Corollary 8.** For a standard two-process parallel model, (i) if the cumulative hazard function $H(t)$ is concave or linear, then $S^2(t) - \frac{S(T_a+t)}{S(T_a)} < 0$; (ii) if $H(t)$ is convex, then the sign of $S^2(t) - \frac{S(T_a+t)}{S(T_a)}$ is uncertain.

The quick interpretation is that anything as slow or slower than an exponential (Recall that an exponential distribution has a linear $H(t)$), and therefore constant, hazard function will imply stochastically increasing ICTs. If $H(t)$ is convex, indicating an increasing $h(t)$, then more information must be garnered. The upshot is, as intuition may suggest, that standard parallel models appear to tend to produce increasing ICTs, but that inclination can be defeated by dramatically increasing hazard functions.

## Standard Multiple-Process Models

We now proceed to generalize the theorems for standard two-process models to standard multiple-process models. All the theorems, corollaries, and lemmas in this section are new. Suppose there are processes $1, \cdots, n$ in the model. The corresponding processing times $z_1, \cdots, z_n$ are assumed to be iid with the density function $f$. The corresponding distribution function, survival function, hazard function, and the cumulative hazard function are labeled as $F$, $S$, $h$, and $H$, respectively. At this point, we do not have a completely analytic proof for arbitrary *n*. However, we will write down the general expressions and then pick an arbitrary value of *n* with which to perform the pertinent computations.

### Dependence of Total Completion Times, General *n*

There are $n!$ ways to arrange the $n$ processes in the standard *n*-process serial models. If we were to assay the behavior of models where the identity or location of a process was associated with distinct probability distributions, that would necessitate a more complex notation. However, due to the assumption of iid random variables, we can invoke a much-simplified set of symbols (see Townsend & Ashby, 1983, Chapter 15; Townsend, Wenger & Houpt, 2018; or Houpt, Townsend & Jefferson, 2017, for the more general situation).

Thus, Figure 4 presents two particular cases where processes $1, \cdots, j$ are processed earlier than processes $j+1, \cdots, n$ (Case I) and processes $1, \cdots, j$ are processed later than processes $j+1, \cdots, n$ (Case II). According to the earlier definition, $\mathbb{T}_j$ is the time that is consumed from the onset of processing to the moment that process $j$ is complete, where $1 \leq j \leq n$. $\mathbb{S}_j$ is the time spent from the onset of processing to the completion of stage $j$. Notice that for generality, process *j* can be completed at any stage. Therefore $\mathbb{T}_j$ is not necessarily equal to $\mathbb{S}_j$. Specifically, following case II of Figure 4, the TCT of item *n* is $\mathbb{T}_n = z_{j+1} + \cdots + z_n$ but the TCT for stage *n* is $\mathbb{S}_n = z_{j+1} + \cdots + z_n + z_1 + \cdots + z_j$, where $z_i, i \in \{1,2,\cdots,j,j+1\cdots,n\}$ is the processing time for item *i*.

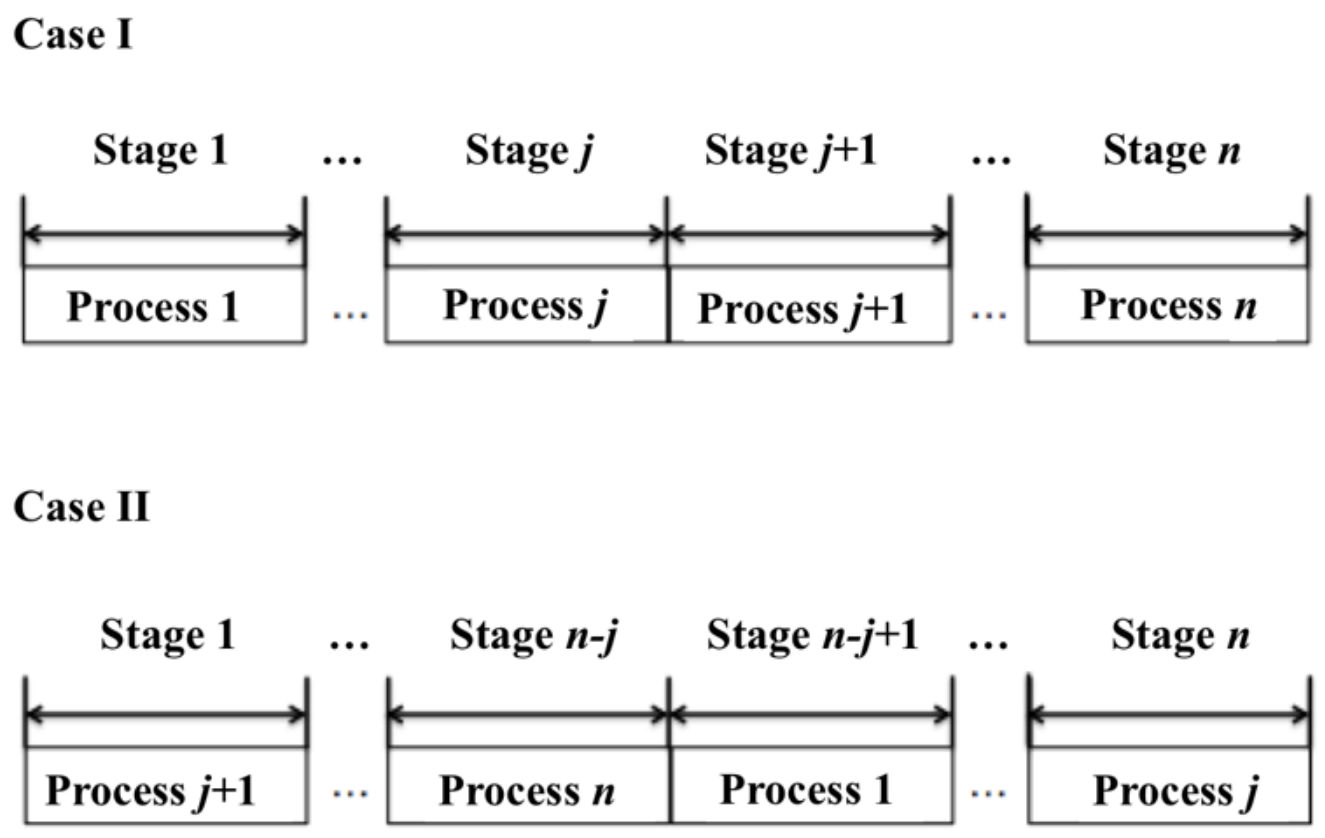

Figure 4. Examples of possible serial arrangements for *n* processes.

Similarly, as for the two-process standard models, we aim to investigate the joint distribution of $\mathbb{T}_{j+1}, \ldots, \mathbb{T}_n$ conditional on $\mathbb{T}_1, \ldots, \mathbb{T}_j$ versus the unconditional distribution function of $\mathbb{T}_{j+1}, \ldots, \mathbb{T}_n$. That is

$$P(\mathbb{T}_{j+1} \leq \tau, \ldots, \mathbb{T}_n \leq \tau \mid \mathbb{T}_1 \leq \tau, \ldots, \mathbb{T}_j \leq \tau) \\ -P\left(\mathbb{T}_{j+1} \leq \tau, \ldots, \mathbb{T}_n \leq \tau\right). \tag{2}$$

Now, if $\mathbb{T}_{j+1}, \ldots, \mathbb{T}_n$ are finished before $\mathbb{T}_1, \ldots, \mathbb{T}_j$, since each of $\mathbb{T}_1, \ldots, \mathbb{T}_j$ includes the amount of $max\ (\mathbb{T}_{j+1}, \ldots, \mathbb{T}_n)$, the probability that $\mathbb{T}_{j+1}, \ldots, \mathbb{T}_n$ are less than $\tau$, given that each of $\mathbb{T}_1, \ldots, \mathbb{T}_j$ already < $\tau$, must be 1. If $\mathbb{T}_1, \ldots, \mathbb{T}_j$ are done first by time $\tau$, the probability that $\mathbb{T}_{j+1}, \ldots, \mathbb{T}_n$ also is finished by $\tau$ is greater than its marginal probability. Thus, this suggests that the TCT in a *n*-process serial model should be positively dependent. We will learn that this reasoning is incorrect without restriction on the processing order but correct when only one processing order is allowed.

At this point, we do not have a completely analytic proof for arbitrary *n*. However, we will write down the general expressions and then pick an arbitrary value of *n* with which to perform the pertinent computations. Theorem 9 is proved based on simulating the function derived from (2).

**Theorem 9.** For a standard *n*-process serial model,

$$P\left(\mathbb{T}_{j+1} \leq \tau, \dots, \mathbb{T}_n \leq \tau \mid \mathbb{T}_1 \leq \tau, \dots, \mathbb{T}_j \leq \tau\right) - P\left(\mathbb{T}_{j+1} \leq \tau, \dots, \mathbb{T}_n \leq \tau\right)$$

can be either positive or nonpositive for $\tau > 0$.

**Proof**. First note that

$$P\left(\mathbb{T}_{j+1} \leq \tau, \dots, \mathbb{T}_n \leq \tau \mid \mathbb{T}_1 \leq \tau, \dots, \mathbb{T}_j \leq \tau\right) = \frac{P\left(\mathbb{T}_{j+1} \leq \tau, \dots, \mathbb{T}_n \leq \tau, \mathbb{T}_1 \leq \tau, \dots, \mathbb{T}_j \leq \tau\right)}{P\left(\mathbb{T}_1 \leq \tau, \dots, \mathbb{T}_j \leq \tau\right)}.$$

We have

$$\begin{aligned} &P\left(\mathbb{T}_{j+1} \leq \tau, \dots, \mathbb{T}_n \leq \tau, \mathbb{T}_1 \leq \tau, \dots, \mathbb{T}_j \leq \tau\right) \\ &= P\left[max\left(\mathbb{T}_{j+1}, \dots, \mathbb{T}_n, \mathbb{T}_1, \dots, \mathbb{T}_j\right) \leq \tau\right] \\ &= P(z_1 + \dots + z_n \leq \tau) \\ &= \int_0^{\tau} \cdots \int_0^{t_3} \int_0^{t_2} f(\tau_n - \tau_{n-1}) \cdots f(\tau_2 - \tau_1) f(\tau_1) d\tau_1 d\tau_2 \cdots d\tau_n \\ &= f(\tau)^{*(n-1)} * F(\tau), \end{aligned}$$

where $f(\tau)^{*m} = \underbrace{f(\tau) * \dots * f(\tau)}_{mf(\tau)'s}$.

Now, we need to compute $P\left(\mathbb{T}_1 \leq \tau, \dots, \mathbb{T}_j \leq \tau\right)$. Since there is no restriction on the order of processing, it is possible to finish *j* processes in the *j*-th stage and any stage after that. We observe that

$$max\left(\mathbb{T}_1, \dots, \mathbb{T}_j\right) \in \left\{\mathbb{S}_j, \mathbb{S}_{j+1}, \cdots, \mathbb{S}_n\right\}.$$

We denote the probability that processes 1 to *j* finish in stage *i* as $p_i$, such that

$$p_i = P\left[max\left(\mathbb{T}_1, \dots, \mathbb{T}_j\right) = \mathbb{S}_i\right],$$

where $i \in \{j, j+1, \cdots, n-1\}$ and the probability that processes 1 to *j* finish in stage *n* as *q*, such that

$$q = P\left[max\left(\mathbb{T}_1, \dots, \mathbb{T}_j\right) = \mathbb{S}_n\right].$$

It is apparent that $p_j + p_{j+1} + \dots + p_{n-1} + q = 1$. The cumulative distribution function of the TCT for the *i*-th stage is

$$\begin{aligned} P\left(\mathbb{S}_i \leq \tau\right) &= \int_0^{\tau} \cdots \int_0^{t_3} \int_0^{t_2} f(\tau_i - \tau_{i-1}) \cdots f(\tau_2 - \tau_1) f(\tau_1) d\tau_1 d\tau_2 \cdots d\tau_i \\ &= f(\tau)^{*(i-1)} * F(\tau) \end{aligned}$$

and for the last stage is

$$\begin{aligned} &P\left(\mathbb{S}_n \leq \tau\right) = P(t_1 + \dots + t_n \leq \tau) \\ &= \int_0^{\tau} \cdots \int_0^{t_3} \int_0^{t_2} f(\tau_n - \tau_{n-1}) \cdots f(\tau_2 - \tau_1) f(\tau_1) d\tau_1 d\tau_2 \cdots d\tau_n \\ &= f(\tau)^{*(n-1)} * F(\tau). \end{aligned}$$

Therefore $P\left(\mathbb{T}_1 \leq \tau, \dots, \mathbb{T}_j \leq \tau\right)$ can be written as

$$
\begin{aligned}
& P\big(\mathbb{T}_1 \le \tau, \dots, \mathbb{T}_j \le \tau\big) \\
&= P\big[max\big(\mathbb{T}_1, \dots, \mathbb{T}_j\big) \le \tau\big] \\
&= p_j P\ (\ \mathbb{S}_j \le \tau\ ) + \cdots + p_{n-1} P\ (\ \mathbb{S}_{n-1} \le \tau\ ) + qP\ (\ \mathbb{S}_n \le \tau\ ) \\
&= p_j\big[f(\tau)^{*(j-1)} * F(\tau)\big] + \cdots + p_{n-1}\big[f(\tau)^{*(n-2)} * F(\tau)\big] + q\big[f(\tau)^{*(n-1)} * F(\tau)\big].
\end{aligned}
$$

Next, we want to compute $P\big(\mathbb{T}_{j+1} \le \tau, \dots, \mathbb{T}_n \le \tau\big)$. There are *n* - *j* processes to be finished. They can be fully executed in the (*n* – *j*)-th stage and any stage after that. Therefore, we have

$$max\big(\mathbb{T}_{j+1}, \dots, \mathbb{T}_n\big) \in \{\mathbb{S}_{n-j}, \mathbb{S}_{n-j+1}, \cdots, \mathbb{S}_n\}.$$

Here we denote the probability that process *j* + 1 to process *n* finish in stage *k* as $q_k$, such that

$$q_k = P\big[max\big(\mathbb{T}_{j+1}, \dots, \mathbb{T}_n\big) = \mathbb{S}_k\big],$$

where $k \in \{n-j, n-j+1, \cdots, n-1\}$. Recall that *q* denotes the probability that process 1 to process *j* complete in the last stage. It is equivalent to the probability that process *j* + 1 to process *n* finish before stage *n*. Therefore, it follows that

$$q_{n-j} + q_{n-j+1} + \cdots + q_{n-1} = q$$

and following the similar reasoning,

$$P\big[max\big(\mathbb{T}_{j+1}, \dots, \mathbb{T}_n\big) = \mathbb{S}_n\big] = p_j + p_{j+1} + \cdots + p_{n-1}.$$

Combining the above two equations together, we have

$$p_j + p_{j+1} + \cdots + p_{n-1} + q_{n-j} + q_{n-j+1} + \cdots + q_{n-1} = 1.$$

Thus,

$$
\begin{aligned}
& P\big(\mathbb{T}_{j+1} \le \tau, \dots, \mathbb{T}_n \le \tau\big) \\
&= P\big[max\big(\mathbb{T}_{j+1}, \dots, \mathbb{T}_n\big) \le \tau\big] \\
&= q_{n-j} P\big(\ \mathbb{S}_{n-j} \le \tau\ \big) + \cdots + q_{n-1} P\big(\ \mathbb{S}_{n-1} \le \tau\ \big) + (p_j + p_{j+1} + \cdots \\
&\qquad + p_{n-1}) P\big(\ \mathbb{S}_n \le \tau\ \big) \\
&= q_{n-j}\big[f(\tau)^{*(n-j-1)} * F(\tau)\big] + \cdots + q_{n-1}\big[f(\tau)^{*(n-2)} * F(\tau)\big] \\
&\qquad + \big(p_j + p_{j+1} + \cdots + p_{n-1}\big)\big[f(\tau)^{*(n-1)} * F(\tau)\big].
\end{aligned}
$$

Therefore, we arrive at the complicated expression

$$
\begin{aligned}
& P\ (\ \mathbb{T}_{j+1} \le \tau, \dots, \mathbb{T}_n \le \tau \mid \mathbb{T}_1 \le \tau, \dots, \mathbb{T}_j \le \tau\ ) - P\ \big(\mathbb{T}_{j+1} \le \tau, \dots, \mathbb{T}_n \le \tau\big) \\
&= \frac{f(\tau)^{*(n-1)} * F(\tau)}{p_j\big[f(\tau)^{*(j-1)} * F(\tau)\big] + \cdots + p_{n-1}\big[f(\tau)^{*(n-2)} * F(\tau)\big] + q\big[f(\tau)^{*(n-1)} * F(\tau)\big]} - \big\{q_{n-j}\big[f(\tau)^{*(n-j-1)} * F(\tau)\big] + \\
&\qquad \cdots + q_{n-1}\big[f(\tau)^{*(n-2)} * F(\tau)\big] + \big(p_j + p_{j+1} + \cdots + p_{n-1}\big)\big[f(\tau)^{*(n-1)} * \\
&\qquad F(\tau)\big]\big\}.
\end{aligned}
\tag{3}
$$

We are now in a position to investigate the sign of (3). (3) takes account of all the possible permutations for the $n$ processes. If we consider two permutations as illustrated in Figure 4, then

$$p_{j+1} = \cdots = p_{n-1} = q_{n-j+1} = \cdots = q_{n-1} = 0.$$

If we can prove (3) can be both positive and nonpositive for the two permutations, the statement of this theorem will be proven for a given value of *n*. At this point we see that

$$(3) = \frac{f(\tau)^{*(n-1)} * F(\tau)}{p_j\left[f(\tau)^{*(j-1)} * F(\tau)\right] + q_{n-j}\left[f(\tau)^{*(n-1)} * F(\tau)\right]} - \left\{q_{n-j}\left[f(\tau)^{*(n-j-1)} * F(\tau)\right] + p_j\left[f(\tau)^{*(n-1)} * F(\tau)\right]\right\}$$

$$= R''\left\{1 - p_j q_{n-j}\left\{\left[f(\tau)^{*(n-1)} * F(\tau)\right]^{\frac{1}{2}} - \frac{\left[f(\tau)^{*(n-j-1)} * F(\tau)\right]^{\frac{1}{2}}\left[f(\tau)^{*(j-1)} * F(\tau)\right]^{\frac{1}{2}}}{\left[f(\tau)^{*(n-1)} * F(\tau)\right]^{\frac{1}{2}}}\right\}^2 + p_j q_{n-j}\left\{\left[f(\tau)^{*(j-1)} * F(\tau)\right]^{\frac{1}{2}} - \left[f(\tau)^{*(n-j-1)} * F(\tau)\right]^{\frac{1}{2}}\right\}^2 - p_j\left[f(\tau)^{*(j-1)} * F(\tau)\right] - q_{n-j}\left[f(\tau)^{*(n-j-1)} * F(\tau)\right]\right\},$$

(4)

where $R'' = \frac{f(\tau)^{*(n-1)} * F(\tau)}{p_j\left[f(\tau)^{*(j-1)} * F(\tau)\right] + q_{n-j}\left[f(\tau)^{*(n-1)} * F(\tau)\right]}$ and $p_j + q_{n-j} = 1$. If $f(\tau)^{*(j-1)} * F(\tau) = f(\tau)^{*(n-j-1)} * F(\tau)$, that is $j = \frac{n}{2}$, and then (4) is reduced to

$$R''\left\{1 - p_j q_{n-j}\left\{\left[f(\tau)^{*(n-1)} * F(\tau)\right]^{\frac{1}{2}} - \frac{f(\tau)^{*(j-1)} * F(\tau)}{\left[f(\tau)^{*(n-1)} * F(\tau)\right]^{\frac{1}{2}}}\right\}^2 - \left[f(\tau)^{*(j-1)} * F(\tau)\right]\right\}.$$

Since $p_j q_{n-j} \leq \frac{1}{4}$, the above expression is

$$\geq R"\left\{1-\frac{1}{4}\left\{\left[f(\tau)^{*(n-1)}*F(\tau)\right]^{\frac{1}{2}}-\frac{f(\tau)^{*(j-1)}*F(\tau)}{\left[f(\tau)^{*(n-1)}*F(\tau)\right]^{\frac{1}{2}}}\right\}^{2}-\left[f(\tau)^{*(j-1)}*F(\tau)\right]\right\}.$$

(5)

Note that if $p_j = \frac{1}{2}$, the above $\geq$ reduces to $=$. Now, it is requisite to investigate the sign of (5). If it can be both positive and nonpositive, it indicates that (2) can be positive and nonpositive. We have this ordering:

$$0 < f(\tau)^{*(n-1)} * F(\tau) \leq f(\tau)^{*(j-1)} * F(\tau) \leq 1 \leq \frac{f(\tau)^{*(j-1)} * F(\tau)}{\left[f(\tau)^{*(n-1)} * F(\tau)\right]^{\frac{1}{2}}},$$

as $\frac{f(\tau)^{*(j-1)}*F(\tau)}{\left[f(\tau)^{*(n-1)}*F(\tau)\right]^{\frac{1}{2}}} = \left[\frac{f(\tau)^{*(j-1)}*F(\tau)f(\tau)^{*(j-1)}*F(\tau)}{f(\tau)^{*(n-1)}*F(\tau)}\right]^{\frac{1}{2}} = \left[\frac{f(\tau)^{*(j-1)}*F(\tau)f(\tau)^{*(n-j-1)}*F(\tau)}{f(\tau)^{*(n-1)}*F(\tau)}\right]^{\frac{1}{2}} =$
$\left[\frac{P(\mathbb{S}_j \leq \tau)P(\mathbb{S}_{n-j} \leq \tau)}{P(\mathbb{S}_n \leq \tau)}\right]^{\frac{1}{2}} = \left[\frac{P(max\ (\mathbb{S}_j, \mathbb{S}_{n-j}) \leq \tau}{P(\mathbb{S}_n \leq \tau)}\right]^{\frac{1}{2}} \geq 1.$

The sign of (5) was computed using the simulation-based method the same as that was used for *n* = 2 following these steps:

Step 1: generate a random number $\alpha \sim Uniform\ [0,1]$, where $\alpha$ represents $f(\tau)^{*(n-1)} * F(\tau)$.

Step 2: generate a random number $\beta \sim Uniform\ [\alpha, 1]$, where $\beta$ represents $f(\tau)^{*(j-1)} * F(\tau)$.

Step 3: if $\frac{\beta^2}{\alpha} \geq 1$, then compute if the part after $R$" for (5) $> 0$.

It was found that the probability of the part after $R$" for (5) $> 0$ given $\frac{\beta^2}{\alpha} \geq 1$ is 62%. The simulation result indicates that (5) can be both positive or nonpositive. □

We proceed construct several examples that help to clarify Theorem 9. We assume that there are three processes in a standard serial model, whose processing times are iid and labeled as:

$$z_1, z_2, z_3.$$

Of course, one can construct examples that have more than three processes in the models. Here we only discuss standard three-process serial models. Adapting the above expressions, and without loss of generality, one is interested in the signs of the two functions below:

$$P(\mathbb{T}_1 \leq \tau | \mathbb{T}_2 \leq \tau, \mathbb{T}_3 \leq \tau) - P(\mathbb{T}_1 \leq \tau),$$
$$P(\mathbb{T}_1 \leq \tau, \mathbb{T}_2 \leq \tau | \mathbb{T}_3 \leq \tau) - P(\mathbb{T}_1 \leq \tau, \mathbb{T}_2 \leq \tau).$$

We will only discuss the behavior of first of the two above functions. The other function can be examined in an analogous fashion.

According to the earlier definition of $\mathbb{T}$ and $\mathbb{S}$, we see that

$$max(\mathbb{T}_2, \mathbb{T}_3) \in \{\mathbb{S}_2, \mathbb{S}_3\},$$
$$\mathbb{T}_1 \in \{\mathbb{S}_1, \mathbb{S}_2, \mathbb{S}_3\}.$$

We subsequently denote

$$P[max(\mathbb{T}_2, \mathbb{T}_3) = \mathbb{S}_2] = p_2,$$
$$P[\mathbb{T}_1 = \mathbb{S}_1] = q_1,$$
$$P[\mathbb{T}_1 = \mathbb{S}_2] = q_2.$$

Then apparently, we have

$$P[max(\mathbb{T}_2, \mathbb{T}_3) = \mathbb{S}_3] = q_1 + q_2,$$
$$P[\mathbb{T}_1 = \mathbb{S}_3] = p_2,$$

and

$$p_2 + q_1 + q_2 = 1.$$

Consequently,

$$\begin{aligned} &P(\mathbb{T}_2 \le \tau, \mathbb{T}_3 \le \tau) \\ &= P(max\ (\mathbb{T}_2, \mathbb{T}_3) \le \tau) \\ &= p_2 P(\mathbb{S}_2 \le \tau) + (q_1 + q_2) P(\mathbb{S}_3 \le \tau) \\ &= p_2 f(\tau) * F(\tau) + (q_1 + q_2) f(\tau) * f(\tau) * F(\tau), \end{aligned}$$

$$\begin{aligned} &P(\mathbb{T}_1 \le \tau) \\ &= q_1 P(\mathbb{S}_1 \le \tau) + q_2 P(\mathbb{S}_2 \le \tau) + p_2 P(\mathbb{S}_3 \le \tau) \\ &= q_1 F(\tau) + q_2 f(\tau) * F(\tau) + p_2 f(\tau) * f(\tau) * F(\tau). \end{aligned}$$

Therefore,

$$\begin{aligned} &P(\mathbb{T}_1 \le \tau | \mathbb{T}_2 \le \tau, \mathbb{T}_3 \le \tau) - P(\mathbb{T}_1 \le \tau) \\ &= \frac{P(\mathbb{T}_1 \le \tau, \mathbb{T}_2 \le \tau, \mathbb{T}_3 \le \tau)}{P(\mathbb{T}_2 \le \tau, \mathbb{T}_3 \le \tau)} - P(\mathbb{T}_1 \le \tau) \\ &= \frac{f(\tau) * f(\tau) * F(\tau)}{p_2 f(\tau) * F(\tau) + (q_1 + q_2) f(\tau) * f(\tau) * F(\tau)} \\ &\qquad - \left(q_1 F(\tau) + q_2 f(\tau) * F(\tau) + p_2 f(\tau) * f(\tau) * F(\tau)\right). \end{aligned} \tag{6}$$

Let us consider some specific often employed, distributions.

**Weibull distributions.** Let $z_1, z_2, z_3$ be iid and follow the Weibull distribution with the density function

$$f(\tau) = ku(u\tau)^{k-1} exp[-(u\tau)^k],$$

where the parameters $k, u > 0$. We used simulation methods to compute the values of (6) by selecting $p_2 = q_1 = q_2 = \frac{1}{3}$. Here we present plots for (6) by varying the values of $\tau$ and $u$ (Figure 5). We allowed $u$ to vary from .5 to 10 and $\tau$ to vary from 0.01 to 5. Figure 5(a) fixes $k = .5$, Figure 5(b) fixes $k = 1$, where the Weibull distributions reduce to exponential distributions, and Figure 5(c) fixes $k = 1.5$. The three plots for (6) are non-negative.

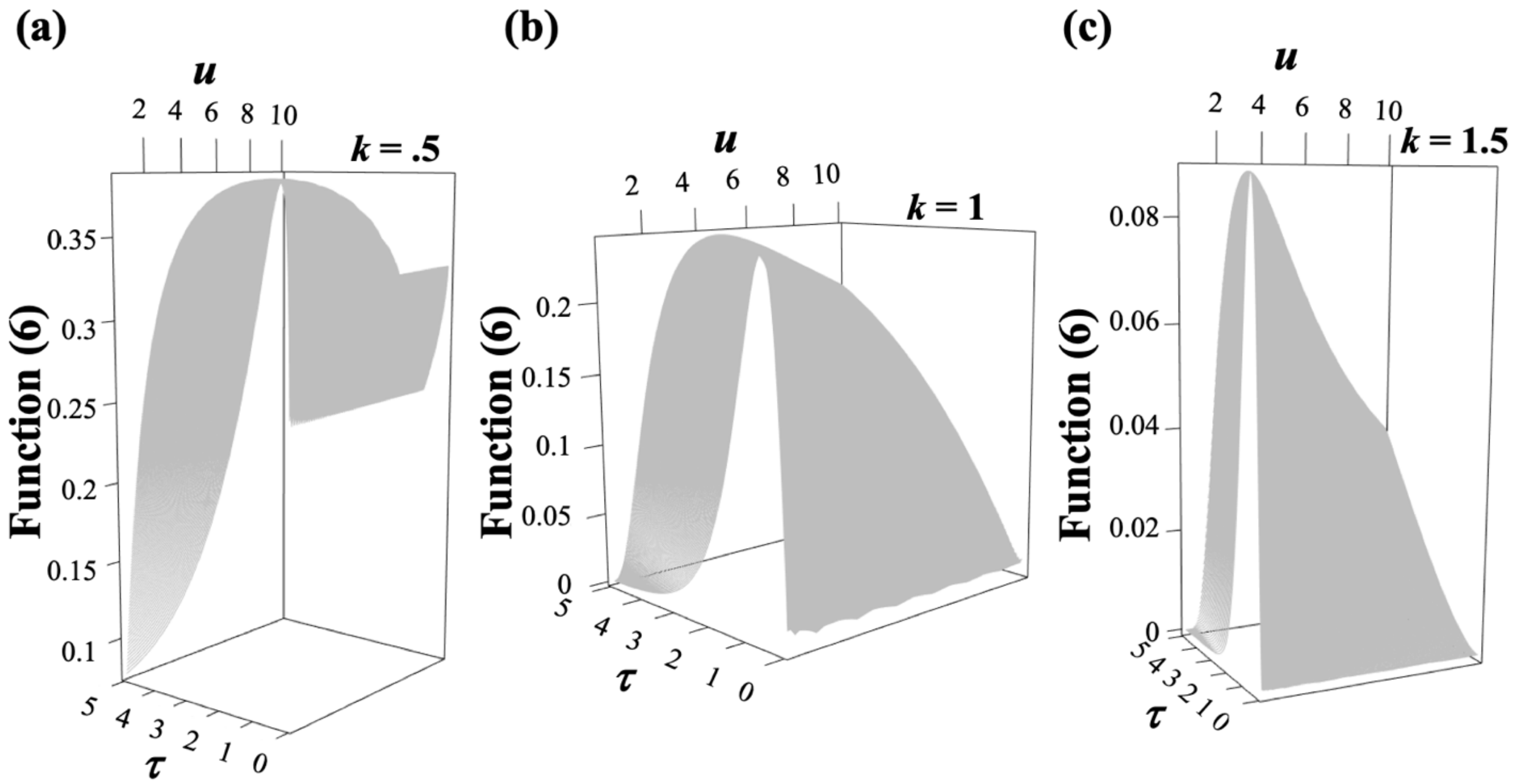

Figure 5. The plots of function (6) for (a) $k = .5$, (b) $k = 1$, and (c) $k = 1.5$. Note that $\tau$ has the arbitrary unit.

Next, we explore the uniform distribution. Although perhaps, unrealistic in fact, it has often been employed to represent the so-called base time distribution, namely the duration which captures the additional unmodeled psychological processes such as early sensory and late motor stages.

**Uniform distributions**. Let $z_1, z_2, z_3$ be iid and follow the uniform distribution

$$z_1, z_2, z_3 \sim Uniform(0, v),$$

where $v > 0$. The corresponding distribution function is

$$F(\tau) = \begin{cases} \frac{\tau}{v}, if\ 0 \leq \tau \leq v \\ 1, otherwise \end{cases}.$$

Also, we have

$$f(\tau) * F(\tau) = \begin{cases} \int_0^\tau \frac{\tau_2}{v^2} d\tau_2 = \frac{\tau^2}{2v^2}, if\ 0 \leq \tau < v, \\ \int_v^\tau \frac{2v - \tau_2}{v^2} d\tau_2 + \frac{1}{2} = \frac{2\tau}{v} - \frac{\tau^2}{2v^2} - 1, if\ v \leq \tau < 2v, \\ 1, if\ 2v \leq \tau. \end{cases}$$

$$f(\tau) * f(\tau) * F(\tau)$$

$$= \begin{cases} \int_0^\tau \frac{\tau_3^2}{2v^3} d\tau_3 = \frac{\tau^3}{6v^3}, if\ 0 \le \tau < v, \\ \int_v^\tau \left( -\frac{\tau_3^2}{v^3} + \frac{3\tau_3}{v^2} - \frac{3}{2v} \right) d\tau_3 + \frac{1}{6} = -\frac{\tau^3}{3v^3} + \frac{3\tau^2}{2v^2} - \frac{3\tau}{2v} + \frac{1}{2}, if\ v \le \tau < 2v, \\ \int_{2v}^\tau \left( \frac{\tau_3^2}{2v^3} - \frac{3\tau_3}{v^2} + \frac{9}{2v} \right) d\tau_3 + \frac{5}{6} = \frac{\tau^3}{6v^3} - \frac{3\tau^2}{2v^2} + \frac{9\tau}{2v} - \frac{7}{2}, if\ 2v \le \tau < 3v, \\ 1, if\ 3v \le \tau. \end{cases}$$

For $3v \le \tau$,

$$\begin{aligned} &P(\mathbb{T}_1 \le \tau | \mathbb{T}_2 \le \tau, \mathbb{T}_3 \le \tau) - P(\mathbb{T}_1 \le \tau) \\ &= \frac{f(\tau) * f(\tau) * F(\tau)}{p_2 f(\tau) * F(\tau) + (q_1 + q_2) f(\tau) * f(\tau) * F(\tau)} \\ &\quad - \big(q_1 F(\tau) + q_2 f(\tau) * F(\tau) + p_2 f(\tau) * f(\tau) * F(\tau)\big) \\ &= \frac{1}{p_2 + (q_1 + q_2)} - (q_1 + q_2 + p_2) \\ &= 0; \end{aligned}$$

for $2v \le \tau < 3v$,

$$\begin{aligned} &P(\mathbb{T}_1 \le \tau | \mathbb{T}_2 \le \tau, \mathbb{T}_3 \le \tau) - P(\mathbb{T}_1 \le \tau) \\ &= \frac{f(\tau) * f(\tau) * F(\tau)}{p_2 f(\tau) * F(\tau) + (q_1 + q_2) f(\tau) * f(\tau) * F(\tau)} \\ &\quad - \big(q_1 F(\tau) + q_2 f(\tau) * F(\tau) + p_2 f(\tau) * f(\tau) * F(\tau)\big) \\ &= \frac{f(\tau) * f(\tau) * F(\tau)}{p_2 + (q_1 + q_2) f(\tau) * f(\tau) * F(\tau)} - \big(q_1 + q_2 + p_2 f(\tau) * f(\tau) * F(\tau)\big) \\ &= \frac{-p_2(1 - p_2)}{p_2 + (q_1 + q_2) f(\tau) * f(\tau) * F(\tau)} \Big[\big(f(\tau) * f(\tau) * F(\tau)\big)^2 - 2 f(\tau) * f(\tau) \\ &\quad * F(\tau) + 1\Big] \\ &\le 0; \end{aligned}$$

for $v \le \tau < 2v$ and $0 \le \tau < v$, we ran simulations to investigate the behavior of (6) and the result is plotted in Figure 6. Again, we fixed $p_2 = q_1 = q_2 = \frac{1}{3}$ and $v = 2$. By observing the plot, we conclude that (6) can, like the preceding example be either positive or nonpositive for uniformly distributed processing times.

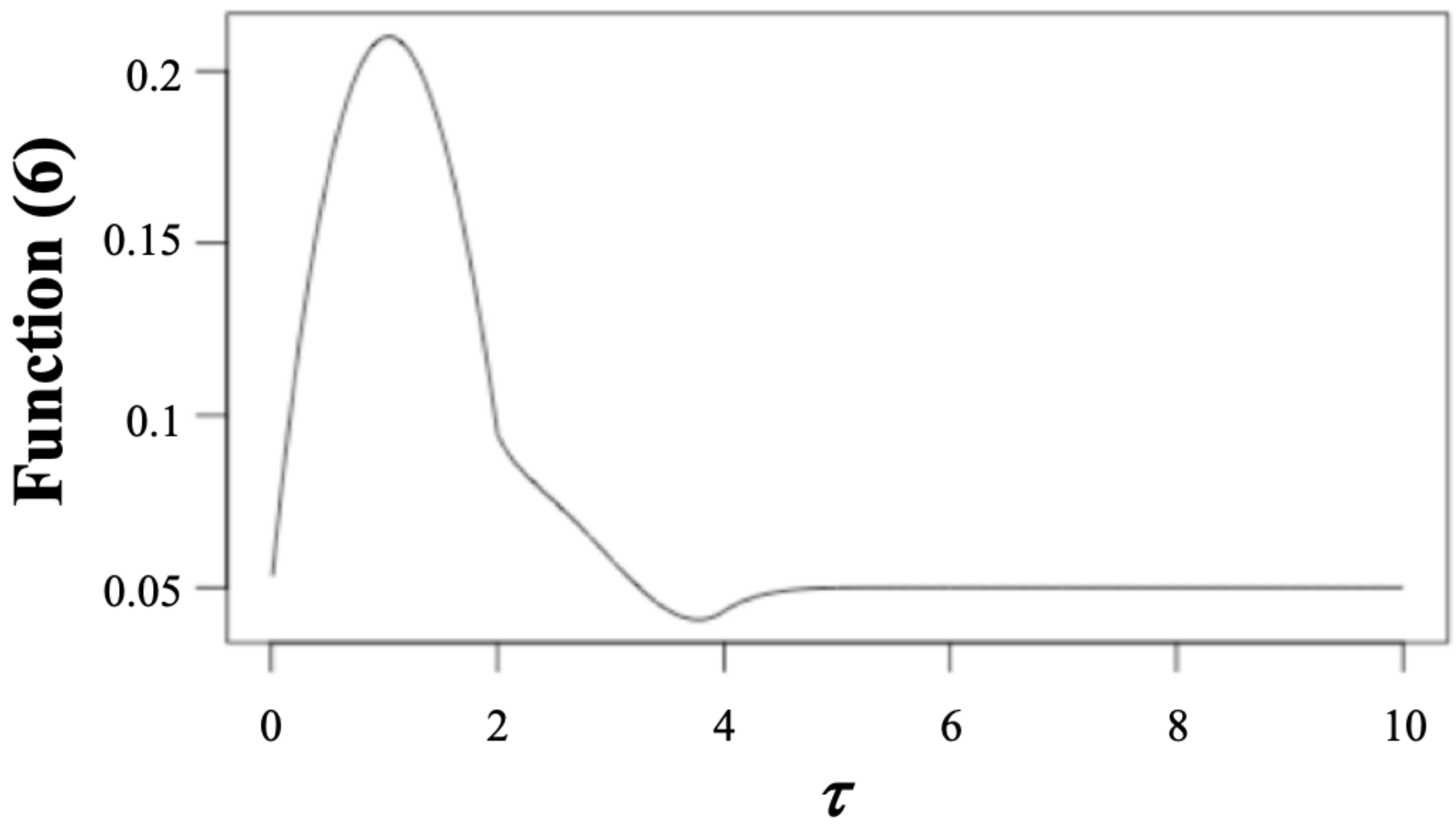

Figure 6. The plot of function (6) for uniformly distributed processing times. Note that $\tau$ has the arbitrary unit.

It is now a good place to demonstrate, for any *n*, that if only one order of processing is allowed, the dependence of total completion times is always nonnegative. This too, generalizes our earlier results for *n* = 2.

**Corollary 10.** For a standard *n*-process serial model, $P(\mathbb{T}_{j+1} \leq \tau, \dots, \mathbb{T}_n \leq \tau \mid \mathbb{T}_1 \leq \tau, \dots, \mathbb{T}_j \leq \tau) - P\left(\mathbb{T}_{j+1} \leq \tau, \dots, \mathbb{T}_n \leq \tau\right)$ is non-negative for $\tau > 0$ if only one permutation is allowed.

**Proof.** According to the derivation in Theorem 9, we have

$$P(\mathbb{T}_{j+1} \leq \tau, \dots, \mathbb{T}_n \leq \tau \mid \mathbb{T}_1 \leq \tau, \dots, \mathbb{T}_j \leq \tau) - P\left(\mathbb{T}_{j+1} \leq \tau, \dots, \mathbb{T}_n \leq \tau\right)$$

$$= \frac{f(\tau)^{*(n-1)} * F(\tau)}{p_j\left[f(\tau)^{*(j-1)} * F(\tau)\right] + \cdots + p_{n-1}\left[f(\tau)^{*(n-2)} * F(\tau)\right] + q\left[f(\tau)^{*(n-1)} * F(\tau)\right]}$$
$$- \left\{q_{n-j}\left[f(\tau)^{*(n-j-1)} * F(\tau)\right] + \cdots + q_{n-1}\left[f(\tau)^{*(n-2)} * F(\tau)\right] + \left(p_j + p_{j+1} + \cdots + p_{n-1}\right)\left[f(\tau)^{*(n-1)} * F(\tau)\right]\right\}.$$

If one permutation is allowed, one and only one element in the set

$$\left\{p_j, p_{j+1}, \cdots, p_{n-1}, q_{n-j}, q_{n-j+1}, \cdots, q_{n-1}\right\}$$

is 1 and the other elements are 0s. So, the above equation can be written as

$$= R\Big\{f(\tau)^{*(n-1)} * F(\tau)$$
$$- \Big\{q_{n-j}\left[f(\tau)^{*(n-j-1)} * F(\tau)\right] + \cdots + q_{n-1}\left[f(\tau)^{*(n-2)} * F(\tau)\right]$$
$$+ (p_j + p_{j+1} + \cdots + p_{n-1})\left[f(\tau)^{*(n-1)} * F(\tau)\right]\Big\}\Big\{p_j\left[f(\tau)^{*(j-1)} * F(\tau)\right] + \cdots$$
$$+ p_{n-1}\left[f(\tau)^{*(n-2)} * F(\tau)\right] + q\left[f(\tau)^{*(n-1)} * F(\tau)\right]\Big\}\Big\}$$
$$= R\left\{f(\tau)^{*(n-1)} * F(\tau) - f(\tau)^{*(l-1)} * F(\tau)\left[f(\tau)^{*(n-1)} * F(\tau)\right]\right\} \geq 0$$

for $\tau > 0$, where

$$R = \frac{1}{p_j\left[f(\tau)^{*(j-1)} * F(\tau)\right] + \cdots + p_{n-1}\left[f(\tau)^{*(n-2)} * F(\tau)\right] + q\left[f(\tau)^{*(n-1)} * F(\tau)\right]}$$

and

$$l \in \{min(j, n-j), \ldots, n-1\}.$$

□

Finally, we also can show that the coarser statistic, the covariance, between any two, of the subprocesses, is always positive.

**Lemma 11.** For a standard *n*-process serial model, again assuming a single processing order,

$Cov(\mathbb{T}_j, \mathbb{T}_l) = \sum_{i=k}^{n}\sum_{z=k}^{n}\sum_{k=1}^{n} m_i w_z Var(T_k)$, $1 \leq j, l \leq n$, and $l \neq j$.

Here $\mathbb{T}_j = \sum_{i=1}^{n} m_i \mathbb{S}_i$, where $m_i$ is the probability for $\mathbb{T}_j = \mathbb{S}_i$ and $\sum_{i=1}^{n} m_i = 1$. $\mathbb{T}_l = \sum_{z=1}^{n} w_z \mathbb{S}_z$, where $w_z$ is the probability for $\mathbb{T}_l = \mathbb{S}_z$ and $\sum_{z=1}^{n} w_z = 1$.

**Proof.**

$$Cov(\mathbb{T}_j, \mathbb{T}_l) = Cov\left(\sum_{i=1}^{n} m_i \mathbb{S}_i, \sum_{z=1}^{n} w_z \mathbb{S}_z\right)$$
$$= Cov\left(\sum_{i=1}^{n} m_i \sum_{k=1}^{i} T_k, \sum_{z=1}^{n} w_z \sum_{k=1}^{z} T_k\right)$$
$$= Cov\left(\sum_{i=1}^{n}\sum_{k=1}^{i} m_i T_k, \sum_{z=1}^{n}\sum_{k=1}^{z} w_z T_k\right)$$
$$= \sum_{i=k}^{n}\sum_{z=k}^{n}\sum_{k=1}^{n} m_i w_z Var(T_k).$$

□

In standard multiple-process parallel models, by definition, the TCT for a process is the processing time for that channel. That implies that the TCTs for processes $\mathbb{T}_1, \ldots, \mathbb{T}_n$ are identically and independently distributed. Similarly, as for the standard *n*-process serial models, we state the behavior of (2) for the standard *n*-process parallel models, which we know from the prior knowledge, to possess no dependence. We include the statement of these tautologies to contrast the distinct predictions of the standard serial vs. the standard parallel models.

**Theorem 12.** For a standard *n*-process parallel model, $P\ (\ \mathbb{T}_{j+1} \leq\ \tau, \ldots, \mathbb{T}_n \leq \tau \,|\, \mathbb{T}_1 \leq\ \tau, \ldots, \mathbb{T}_j\ \leq\ \tau\ ) - P\left(\mathbb{T}_{j+1} \leq\ \tau, \ldots, \mathbb{T}_n \leq\ \tau\right) = 0$ for $\tau > 0$.

**Proof.** For a standard *n*-process parallel model,

$$P\ (\ \mathbb{T}_{j+1} \leq\ \tau, \ldots, \mathbb{T}_n \leq\ \tau \,|\, \mathbb{T}_1 \leq\ \tau, \ldots, \mathbb{T}_j\ \leq\ \tau\ ) - P\left(\mathbb{T}_{j+1} \leq\ \tau, \ldots, \mathbb{T}_n \leq\ \tau\right)$$

$$= \frac{P\ (\ \mathbb{T}_{j+1} \leq\ \tau, \ldots, \mathbb{T}_n \leq\ \tau\ , \mathbb{T}_1 \leq\ \tau, \ldots, \mathbb{T}_j\ \leq\ \tau\ \ )}{P\left(\mathbb{T}_1 \leq\ \tau, \ldots, \mathbb{T}_j\ \leq\ \tau\ \right)} - P\left(\mathbb{T}_{j+1} \leq\ \tau, \ldots, \mathbb{T}_n \leq\ \tau\right)$$

$$= \frac{P\left(\ \mathbb{T}_{j+1} \leq\ \tau, \ldots, \mathbb{T}_n \leq\ \tau\ \right) P\left(\ \mathbb{T}_1 \leq\ \tau, \ldots, \mathbb{T}_j\ \leq\ \tau\ \ \right)}{P\left(\mathbb{T}_1 \leq\ \tau, \ldots, \mathbb{T}_j\ \leq\ \tau\ \right)} - P\left(\mathbb{T}_{j+1} \leq\ \tau, \ldots, \mathbb{T}_n \leq\ \tau\right)$$

$$= 0.$$ □

And, of course, the covariance is destined to be 0 as well.

**Lemma 13.** For a standard *n*-process parallel model,

$$Cov\left(\mathbb{T}_j, \mathbb{T}_l\right) = 0,$$

where $1 \leq j, l \leq n$, and $l \neq j$.

**Proof.** It is apparent. □

Our investigation of TCTs for general *n* shows that dependence of these in the case of standard serial models can be either positive or nonpositive. The computations of special cases strongly suggest that positive dependencies are much easier to come by, perhaps because of our earlier expressed intuitions in the section for the standard two-process models.

Furthermore, the above derivations suggest that the behavior of standard serial and standard parallel models differ substantively. Function (2) = 0 looks difficult for a distribution for serial class of models to satisfy for all $\tau > 0$. However, it is a functional equation that appears to be quite challenging to solve. On the other hand, the case for a single processing order is concrete and clear: Function (2) has to be 0 for a standard parallel model yet cannot be for the standard serial model with a single order. The strategic issue of model mimicry will be revisited in a subsequent section.

**Dependence of Intercompletion Times, General *n***

We now proceed to investigate the general statistic associated with the actual processing times of standard serial models, that is, the ICTs. Without loss of generalization, we can assume that $z_1 \leq z_2 \leq \cdots \leq z_n$. Recall that in standard *n*-process serial models, the ICTs are independent. Therefore, we only investigate the dependence of ICTs for standard *n*-process parallel models.

The standard *n*-process parallel model can be decomposed into *n* stages (Figure 1(b) can be viewed as a standard *n*-process parallel model). We have the ICTs for the standard *n*-process parallel models:

$$\begin{aligned} T_1 &= \mathbb{T}_1 = z_1, \\ T_2 &= \mathbb{T}_2 - \mathbb{T}_1 = z_2 - z_1, \\ &\cdots, \\ T_n &= \mathbb{T}_n - \mathbb{T}_{n-1} = z_n - z_{n-1}. \end{aligned}$$

We investigate the behavior of the survival function of the canonical sum of ICTs $T_{j+1} + \cdots + T_{n'}, 0 < j + 1 \leq n' \leq n$, conditional on the completion of the earlier stages: $P\left(T_{j+1} + \cdots + T_{n'} > t \middle| z_{n'} > z_j\right)$. Without loss of generality, we assume that process 1 terminates first, process 2 terminates second, …, and process *n* terminates in the last. Lemma 14 shows that the ICTs of standard parallel models tend to increase as a function of the previous ICTs in a strong distributional sense if the hazard function is constant or decreasing.

**Lemma 14.** For a standard *n*-process parallel model, if the hazard function $h$ is non-increasing, then $P\left(T_{j+1} + \cdots + T_{n'} > t \middle| z_{n'} > z_j\right)$ is non-decreasing as $z_j$ is increased.

**Proof.** We have

$$\begin{aligned} &P\left(T_{j+1} + \cdots + T_{n'} > t \middle| z_{n'} > z_j\right) \\ &= P\left(T_1 + \cdots + T_j + T_{j+1} + \cdots + T_{n'} > t + T_1 + \cdots + T_j \middle| z_{n'} > z_j\right) \\ &= P\left(z_{n'} > t + z_j \middle| z_{n'} > z_j\right) \\ &= \frac{S(z_j + t)}{S(z_j)}. \end{aligned}$$

To examine the behavior of this function as $z_j$ changes, one can take the derivative

$$\frac{d}{dz_j}\frac{S(z_j+t)}{S(z_j)} = \frac{-S(z_j)f(z_j+t)+S(z_j+t)f(z_j)}{S^2(z_j)}$$
$$= \frac{S(z_j)S(z_j+t)}{S^2(z_j)}\left[\frac{f(z_j)}{S(z_j)} - \frac{f(z_j+t)}{S(z_j+t)}\right]$$
$$= \frac{S(z_j)S(z_j+t)}{S^2(z_j)}\left[h(z_j) - h(z_j+t)\right].$$

If the hazard function is non-increasing, then $\frac{d}{dz_j}\frac{S(z_j+t)}{S(z_j)} \geq 0$. Consequently, $P(T_{j+1}+\cdots+T_{n'} > t | z_{n'} > z_j)$ is non-decreasing as $z_j$ is increased. □

Thus, happily we again find that, as for *n* = 2, a constant or decreasing hazard function forces non-decreasing sums of ICTs.

Next, as before with *n* = 2, we move on to deliberate on what characterizes the successive ICTs for standard parallel models in the general case of arbitrary *n*. We can write the survival functions for the *n*-process parallel models at each stage conditioned on the completion of the earlier stage(s) in Table 1:

Table 1. Survival functions for the standard *n*-process parallel models at each stage.

| Stage | Survival function |
|---|---|
| Stage 1 | $S^n(t)$ |
| Stage 2 | $\left[\frac{S(z_1+t)}{S(z_1)}\right]^{n-1}$ |
| Stage 3 | $\left[\frac{S(z_2+t)}{S(z_2)}\right]^{n-2}$ |
| , …, | , …, |
| Stage *n* | $\frac{S(z_{n-1}+t)}{S(z_{n-1})}$ |

Let us denote the ratio of the hazard functions

$$\alpha(z_{j-1}+t, z_j+t) = \frac{h(z_j+t)}{h(z_{j-1}+t)}.$$

If the survival function from stage *j* to stage *j* + 1 for every *j* is increasing, this trend is then consistent with the empirical finding that the time interval between two successive responses is increasing as stage number is increasing. Theorem 15 provides under what exact condition the survival function keeps increasing from stage *j* to stage *j* + 1. A generalized form of the result for *n* = 2 again finds that the above ratio of hazard functions controls the behavior of the sequence of ICTs. Corollary 16 states that standard *n*-process parallel models with concave or linear $H(t)$ result in the increasing survival functions across stages.

**Theorem 15.** For a standard *n*-process parallel model, the survivor function from the *j*-th stage to the (*j*+1)-th stage is increasing if $\alpha\left(z_{j-1}+t, z_j+t\right) < \frac{n-j+1}{n-j}$ and is non-increasing otherwise.

**Proof.** Consider stage *j* VS. stage *j*+1:

$$\left[\frac{S\left(t+z_{j-1}\right)}{S\left(z_{j-1}\right)}\right]^{n-j+1} - \left[\frac{S\left(t+z_j\right)}{S\left(z_j\right)}\right]^{n-j}$$
$$= \frac{exp\left[-(n-j+1)H(t+z_{j-1})\right]}{exp\left[-(n-j+1)H(z_{j-1})\right]} - \frac{exp\left[-(n-j)H(t+z_j)\right]}{exp\left[-(n-j)H(z_j)\right]}.$$

The sign of the above equation is the same as

$$-(n-j+1)H\left(t+z_{j-1}\right) - (n-j)H\left(z_j\right) + (n-j)H\left(t+z_j\right) + (n-j+1)H\left(z_{j-1}\right), \tag{7}$$

which is equivalent to

$$-(n-j+1)\int_0^t h\left(z_{j-1}+t\right)dt + (n-j)\int_0^t h\left(z_j+t\right)dt$$
$$= \int_0^t -(n-j+1)h\left(z_{j-1}+t\right) + (n-j)\alpha\left(z_{j-1}+t, z_j+t\right)h\left(z_{j-1}+t\right)dt$$
$$= \int_0^t h\left(z_{j-1}+t\right)\left[-(n-j+1) + (n-j)\,\alpha\left(z_{j-1}+t, z_j+t\right)\right]dt$$

$$\begin{cases} \geq 0, if\ \alpha\left(z_{j-1}+t, z_j+t\right) \geq \frac{n-j+1}{n-j} \\ < 0, if\ \alpha\left(z_{j-1}+t, z_j+t\right) < \frac{n-j+1}{n-j}. \end{cases} \tag{8}$$

□

**Corollary 16.** For a standard *n*-process parallel model, (i) if the cumulative hazard function $H(t)$ is concave or linear, then

$$\left[\frac{S(t+z_{j-1})}{S(z_{j-1})}\right]^{n-j+1} - \left[\frac{S(t+z_j)}{S(z_j)}\right]^{n-j} < 0;$$

(ii) if $H(t)$ is convex, then the sign of $\left[\frac{S(t+z_{j-1})}{S(z_{j-1})}\right]^{n-j+1} - \left[\frac{S(t+z_j)}{S(z_j)}\right]^{n-j}$ is uncertain.

**Proof.** (i) If the cumulative hazard function $H(t)$ is concave or linear, then the hazard function $h(t)$ is a decreasing or constant. Consequently, $\alpha\left(z_{j-1}+t, z_j+t\right) < 1$ or = 1. According to Theorem 15, $\left[\frac{S(t+z_{j-1})}{S(z_{j-1})}\right]^{n-j+1} - \left[\frac{S(t+z_j)}{S(z_j)}\right]^{n-j} < 0$ is obtained. (ii) If $H(t)$ is convex, the hazard function $h(t)$ is increasing. Then we have $\alpha\left(z_{j-1}+t, z_j+t\right) = \frac{h(z_j+t)}{h(z_{j-1}+t)} > 1$. It is uncertain if $\alpha\left(z_{j-1}+t, z_j+t\right) \geq \frac{n-j+1}{n-j}$ or $\alpha\left(z_{j-1}+t, z_j+t\right) < \frac{n-j+1}{n-j}$. So, the sign of $\left[\frac{S(t+z_{j-1})}{S(z_{j-1})}\right]^{n-j+1} - \left[\frac{S(t+z_j)}{S(z_j)}\right]^{n-j}$ is uncertain. □

Here we construct examples to further illustrate the behavior of $\left[\frac{S(t+z_{j-1})}{S(z_{j-1})}\right]^{n-j+1} - \left[\frac{S(t+z_j)}{S(z_j)}\right]^{n-j}$. As before, we learn that standard parallel models are characterized by a vital inclination toward ever-increasing intervals between successive completions. But that tendency can be overridden by extremely increasing hazard functions. These findings are illustrated in the subsequent examples and Figures 7 - 9.

To construct examples for *n* > 2, we assume that there are three processes in standard parallel models, whose processing times are iid and labeled as:

$$z_1, z_2, z_3.$$

According to Theorem 15, the survival function from stage *j* to stage *j* + 1 is increasing if $\alpha(z_{j-1}+t, z_j+t) < \frac{n-j+1}{n-j}$. For $n = 3$, that is

$$\alpha(t, z_1+t) = \frac{h(z_1+t)}{h(t)} < \frac{3-1+1}{3-1} = \frac{3}{2}, if\ j = 1,$$

$$\alpha(t+z_1, t+z_2) = \frac{h(z_2+t)}{h(z_1+t)} < \frac{3-2+1}{3-2} = 2, if\ j = 2.$$

For *j* = 1 and 2, (7) is equivalent to

$$-3H(t) - 2H(z_1) + 2H(t+z_1), (9)$$

and

$$-2H(t+z_1) - H(z_2) + H(t+z_2) + 2H(z_1), (10)$$

respectively.

**Weibull distributions.** Let

$$z_1, z_2, z_3 \sim Weibull(k, u),$$

where $k, u > 0$. The corresponding cumulative hazard function and the hazard function are

$$H(t) = u(ut)^{k-1}t$$

and

$$h(t) = uk(ut)^{k-1}.$$

If $k = 1$, then Weibull distributions reduce to exponential distributions:

$$z_1, z_2, z_3 \sim Exp(u).$$

The cumulative hazard function for the exponential distribution is linear:

$$H(t) = ut.$$

The hazard function is a constant:

$$h(t) = u.$$

It is apparent

$$\alpha(t, z_1 + t) = \frac{h(z_1 + t)}{h(t)} = 1 < \frac{3}{2}, if\ j = 1,$$

$$\alpha(t + z_1, t + z_2) = \frac{h(z_2 + t)}{h(z_1 + t)} = 1 < 2, if\ j = 2.$$

Therefore, the survival function for exponentially distributed processes is increasing from the first stage to the second stage to the third stage, as expected.

If $k < 1$, then the cumulative hazard function is concave and the hazard function is increasing. Hence, the survival is also increasing from stage 1 to stage 2 to stage 3 as

$$\alpha(t, z_1 + t) = \frac{h(z_1 + t)}{h(t)} < 1 < \frac{3}{2}, if\ j = 1,$$

$$\alpha(t + z_1, t + z_2) = \frac{h(z_2 + t)}{h(z_1 + t)} < 1 < 2, if\ j = 2.$$

If $k > 1$, then the cumulative hazard function is convex. We ran simulations to investigate the dynamics of the survival function from stage 1 to stage 2 to stage 3. We present 3d plots (Figure 7) of (9) by varying the values of $t$ and $z_1$ from 0 to 10 and fixing $u = 1$. Figure 7(a) fixes $k = 2$ and Figure 7(b) fixes $k = 4$. We also present 3d plots (Figure 8) of (10) by varying the values of $t$ and $z_2$ from 0 to 10 and fixing $z_1 = 1$ and $u = 1$. Figure 8(a) fixes $k = 2$ and Figure 8(b) fixes $k = 4$. We observe both (9) and (10) can be negative and nonnegative. We conclude that for $k > 1$, the survival function from stage 1 to stage 2 to stage 3 is neither always increasing nor always non-increasing.

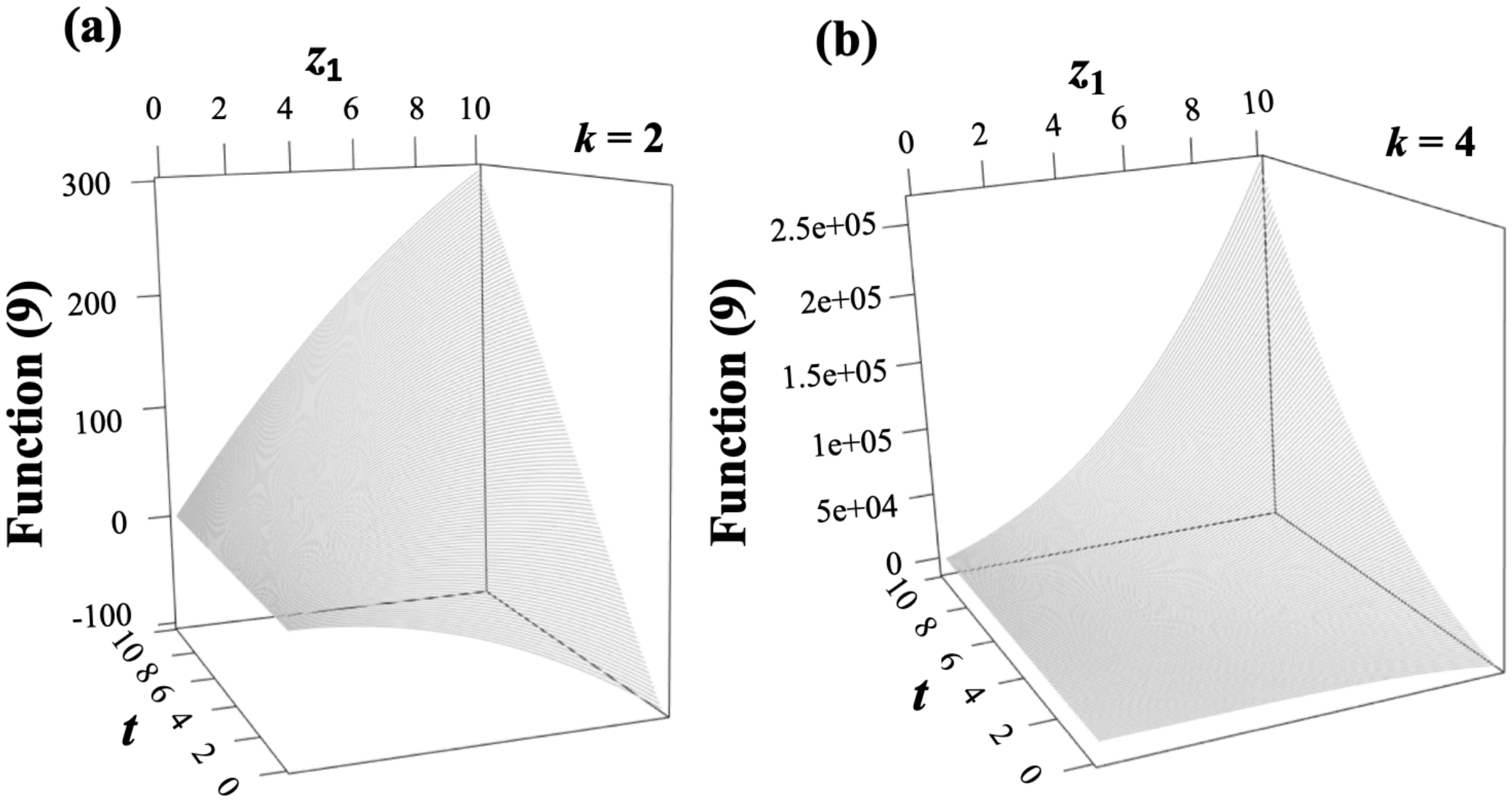

Figure 7. Plots of function (9) for (a) $k = 2$ and (b) $k = 4$. Note that $t$ and $z_1$ have the arbitrary and the same unit.

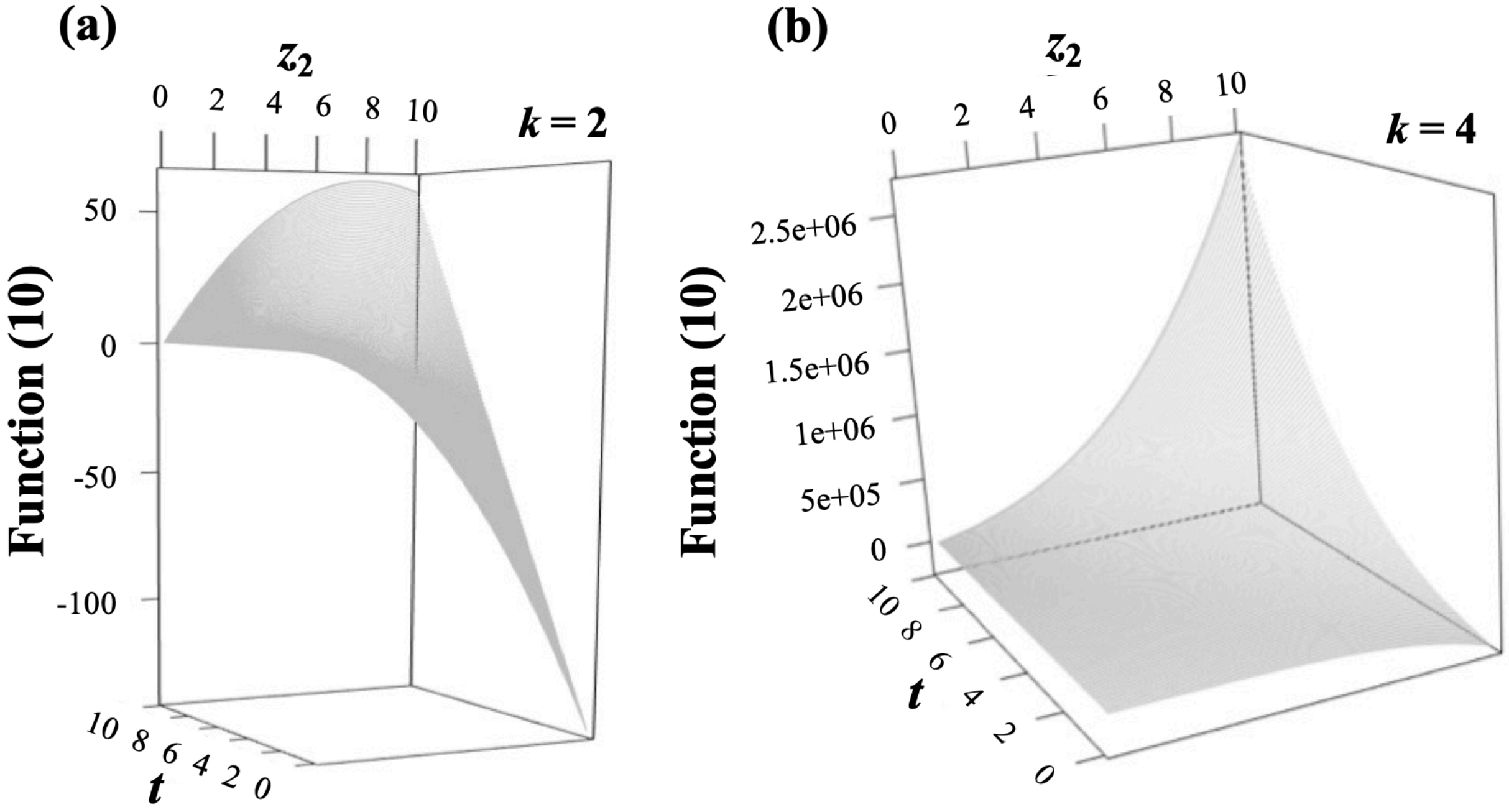

Figure 8. Plots of function (10) for (a) $k = 2$ and (b) $k = 4$. Note that $t$ and $z_2$ have the arbitrary and the same unit.

**Uniform distributions.** Let

$$z_1, z_2, z_3 \sim Uniform(0, v),$$

where $v > 0$. The corresponding cumulative hazard function ($0 \leq t \leq v$) is

convex:

$$H(t) = -ln\,(v - t) + lnv$$

and the hazard function is

$$h(t) = \frac{1}{v - t}.$$

We have

$$\alpha(t, z_1 + t) = \frac{h(z_1 + t)}{h(t)} = \frac{v - t}{v - z_1 - t} \geq 1, if\ j = 1,$$

$$\alpha(t + z_1, t + z_2) = \frac{h(z_2 + t)}{h(z_1 + t)} = \frac{v - z_1 - t}{v - z_2 - t} \geq 1, if\ j = 2.$$

We ran simulations to investigate the dynamics of the survival from stage 1 to stage 2 to stage 3.

We present a 3d plot (Figure 9(a)) of function (9) by varying the values of $t$ and $t_1$ from 0 to 1 and fixing $v = 2$. We also present a 3d plot (Figure 9(b)) of function (10) by varying the values of $t$ and $t_2$ from 0 to 1 and fixing $t_1 = .5$ and $v = 2$. We observe both (9) and (10) can be negative and nonnegative. We conclude that for uniformly distributed processing times, the survival function from stage 1 to stage 2 to stage 3 is neither always increasing nor always non-increasing.

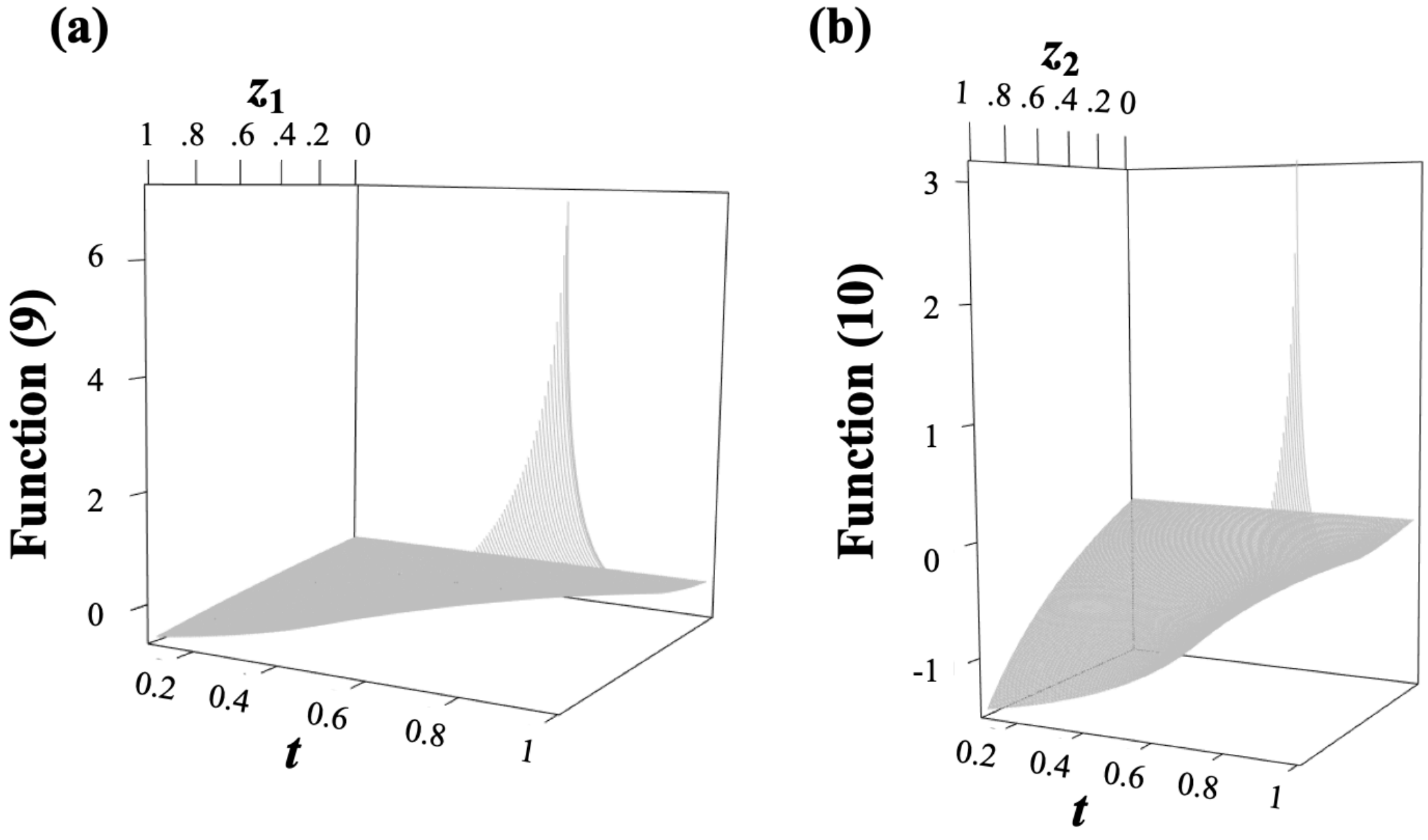

Figure 9. Plots of (a) function (9) and (b) function (10) for uniformly distributed processes. Note that $t$, $z_1$, and $z_2$ have the arbitrary and the same unit.

Analogously to the situation with TCTs, we find that in general ICTs for standard parallel models do not tend to follow the behavior of those for standard serial models which are iid. However, we will proceed to discuss the pivotal issue of mimicry directly.

**Can Standard Serial and Standard Parallel Models Ever Mimic One Another?**

Our very first papers on parallel vs. serial processing uncovered serious issues of mathematical equivalence of these classes in popular experimental paradigms (e.g., Townsend, 1969, 1971, 1972). Later work expanded the classes of distributions subject to mimicry problems (e.g., Townsend, 1976a) as well as pointing to experimental situations and model depictions that might be capable of distinguishing them (e.g., Townsend, 1976b). Yet, those results do not answer the question of model mimicry within the special constraints of standard parallel vs. standard parallel models.

In this realm, the mimicry concern devolves into the inquiry of whether a standard serial model and its assumption of iid ICTs can simultaneously be not only parallel (i.e., equivalent to some arbitrary parallel model) but also produce iid TCTs. The opposite way to pose the question is whether a standard parallel model can be equivalent not only to some arbitrary serial model but to one with iid ICTs. But, observe that this question is actually symmetric because it simply asks whether there is an intersection of the two classes of models that includes the conjunction of iid ICTs and iid TCTs.

We first inspect function (3) in this paper or its special case for *n* = 2 in Zhang, Liu & Townsend (2018) and view either one as a functional equation when the potential inequality is set identically equal to 0. Subjectively, and as intimated earlier, it does not appear that any continuous family of probability densities can satisfy this equation for all times $\tau$. However, we currently have no proof of that.

Hence, we move on to Equation (8) and check if $\alpha\left(z_{j-1}+t, z_j+t\right) \equiv \frac{n-j+1}{n-j}$ for all the values of $z_{j-1}>0, z_j>0, t>0$. We recount the definition of $\alpha$ here that $\alpha\left(z_{j-1}+t, z_j+t\right)=\frac{h\left(z_j+t\right)}{h\left(z_{j-1}+t\right)}$. If $\alpha\left(z_{j-1}+t, z_j+t\right) \equiv \frac{n-j+1}{n-j}$ holds, it indicates standard parallel models can produce iid ICTs, which is a fundamental property of standard serial models. A straightforward proof that no standard parallel model can yield this identity follows.

**Lemma 17.** For a standard *n*-process parallel model, $\alpha\left(z_{j-1}+t, z_j+t\right) \equiv \frac{n-j+1}{n-j}$ cannot hold for all the values of $z_{j-1}>0, z_j>0, t>0$.

**Proof.** Assuming the above equation holds, then we shall have

$$\frac{h\left(z_j + t + (z_j - z_{j-1})\right)}{h\left(z_{j-1} + t + (z_j - z_{j-1})\right)} = \frac{n-j+1}{n-j}$$

$$\frac{h\left(2z_j + t - z_{j-1}\right)}{h\left(z_j + t\right)} = \frac{n-j+1}{n-j}$$

$$\frac{h\left(2z_j + t - z_{j-1}\right)}{\frac{n-j+1}{n-j} h\left(z_{j-1} + t\right)} = \frac{n-j+1}{n-j}$$

$$\frac{h\left(2z_j + t - z_{j-1}\right)}{h\left(z_{j-1} + t\right)} = (\frac{n-j+1}{n-j})^2$$

$$\frac{h\left((2z_j - z_{j-1}) + t\right)}{h\left(z_{j-1} + t\right)} = (\frac{n-j+1}{n-j})^2$$

Contradiction! So $\alpha\left(z_{j-1} + t, z_j + t\right) \equiv \frac{n-j+1}{n-j}$ cannot hold for all the values of $z_{j-1} > 0, z_j > 0, t > 0$. □

Thus, we discern that there exist no standard parallel models can produce iid ICTs. Equivalently, the intersection of the standard serial and standard parallel models is the empty set.

Another similar but perhaps even more intuitive proof of "no intersection of model classes" is the following. Let us observe Lemma 14, if standard parallel models can mimic standard serial models, $P\left(T_{j+1} + \cdots + T_{n'} > t \middle| z_{n'} > z_j\right)$ is invariant across values of $z_j$ and so must have 0 as the only acceptable value of its derivative with respect to $z_j$. That is $\frac{S(z_j)S(z_j+t)}{S^2(z_j)}\left[h(z_j) - h(z_j + t)\right] = 0$. Now, the only continuous hazard function obeying this condition belongs to the exponential family. However, we already know that standard parallel models with exponential distributions possess increasing ICTs (Recall McGill's model in the parallel model's representation) rather than the required independently and identically distributed ones of standard serial models. Hence, once again we find that there is only a null intersection of standard serial and standard parallel models.

We thus find an agreeable consequence of this result: Despite the thorny challenge of model mimicry even between the diametrically opposed types of mental architecture, serial vs. parallel, we discover that the prototypical standard serial and standard parallel models do not mimic one another. Therefore, it makes theoretical and empirical sense to employ such critical statistics as ICT and TCT to test between them in psychological environments.

### Implementation of the Present Theoretical Results in the Laboratory

It is certainly true that both TCTs as well as ICTs yield important information about underlying information processing systems. Nevertheless, when we move to consider how the present findings might be utilized to investigate psychological phenomena, we discover a curious asymmetry. As already observed, ICTs have been studied, though not voluminously, in the realm of free recall. On the other hand, TCTs have not. Apparently, no one has thought of testing the stochastic independence of TCTs in something like a free recall design. It is certainly possible to do and should, in fact, be done.

Actually, in many experimental milieus where processing issues such as architecture, stopping rule, capacity and so on, have been studied, ICTs and TCTs are fairly invisible. But this could change as researchers begin to explore more complex cognitive realms with tools like SFT.

It is also fact that stochastic independence is often more readily assessed when response frequencies, rather than response times, are the primary dependent, observable variable. As we continue to unify approaches which have previously been dominated by response times, such as SFT (e.g., Townsend & Nozawa, 1995) with ones more attached to response frequencies, such as General Recognition Theory (Ashby & Townsend, 1986), our theoretical and statistical power is sure to grow.

Now let us turn to a bit more detailed account of how, at least, our ICT results might relate to the classical free recall paradigm. Therefore, we consider an experimental paradigm in which the subjects recall words from a previously learned list. In the event that the more probable type of standard parallel processing is present we expect a positive association between the magnitude of ICTs for a word and the position of its recall.

Standard serial models cannot interpret this result as we have discussed earlier. For the recall experiment that we propose currently, standard parallel models can readily account for these data. Recall that the TCTs are iid in a standard parallel model. The experimenter records the TCT of each recalled word, which are denoted as $\mathbb{T}_1, \cdots, \mathbb{T}_n$. Let us assume each TCT follows the Weibull distribution:

$$f(\mathbb{T}_j) = ku(u\mathbb{T}_j)^{k-1} exp\left[-(u\mathbb{T}_j)^k\right],$$

where $j \in \{1, \cdots, n\}$. The likelihood function, in this case identical to the joint density function, for a standard parallel model can be written as

$$L = f(\mathbb{T}_1) f(\mathbb{T}_2) \cdots f(\mathbb{T}_n).$$

One can use maximum likelihood method to estimate the parameters $k$ and $u$ for the Weibull distribution. We expect that the estimated value of $k$ is not greater than 1, which is consistent with the prediction of Theorem 15 and Corollary 16.

## Summary and Conclusions

With the purpose of rendering this final section independently readable, we drop our previous acronyms.

Our work here differentiates and characterizes the standard multiple-process serial models and the standard multiple-process parallel models by investigating the behavior of (conditional) distributions of the total completion times and survival functions of intercompletion times without assuming any particular forms for the distributions of processing times. We implement this program through mathematical proofs and computational methods. Although the proofs are more complex than with $n$ fixed at $n = 2$, it pleasantly turned out that the major conclusions are in line with the simpler cases.

Thus, we found that for the standard multiple-process serial models and allowing multiple processing orders, positive dependence between the total completion times may or may not hold if no specific distributional functions are imposed on the processing times. In other words, the conditional probability that processes $j + 1, \ldots, n$ are completed before some time $\tau$ given processes $1, \ldots, j$ have already been completed by this time can be greater or not greater than the unconditional (i.e., marginal) probability that processes $j + 1, \ldots, n$ are completed by time $\tau$**.** Interestingly, the prediction for fixed order serial processing reveals that, unlike the situation with mixtures or orders, standard serial models with a fixed processing order are associated with a positive dependence (e.g., covariance > 0) among total completion times.

By contrast, and per definition, for the standard multiple-process parallel models, the total completion times are independent in the sense that the conditional probability that processes $j + 1, \ldots, n$ are completed before some time $\tau$ given processes $1, \ldots, j$ have already been completed by this time is equal to the unconditional probability that processes $j + 1, \ldots, n$ are completed by time $\tau$. According to the different nature of process dependence, one can distinguish a standard multiple-process serial model from a standard multiple-process parallel model.

Moving on to exploration of what happens in standard parallel models with regard to their intercompletion times, we learn that if the hazard function for the processing times is non-increasing, the later stages tend to be successively longer as a function of the magnitude of the earlier intercompletion time, just as when $n = 2$. Moreover, we discover that in standard multiple-process parallel models, the survival functions of the intercompletion times for the later stages (from stage $j + 1$ to stage $n'$) conditional on the completion of all the earlier stages is non-decreasing as the processing time for $j$-th process is increasing if the hazard function of processing times is non-increasing.

We further find that the survival function of intercompletion time(s) from stage $j$ to stage $j + 1$ is increasing when the ratio of hazard functions is smaller than $\frac{n-j+1}{n-j}$. For all such parallel models the empirical finding that the intercompletion time grows stochastically with the growth of the number of recalled words is accommodated. Finally, if the cumulative hazard function is concave or linear, the survival function from stage $j$ to stage $j + 1$ is increasing.

A limitation of the above theoretical findings is the iid axiom. On the one hand, there is no reason to suspect that as long as a single family of distributions say general gamma, Weibull, Wald, ex-Gaussian and so on, is assumed, that item or position differences should lead to qualitatively different distribution parameters. However, it would be nice to actually demonstrate that fact, as Vorberg and Ulrich (1987) did in the case of the non-iid exponential family. We look forward to a much more rigorous and precise treatment of free recall and related data with the current theory and methodology.

The greatest need appears to be more relevant experimentation. Paradigms associated with free recall as in Bousfield and Sedgewick (e.g., 1944) or Rohror and Wixted (1994) are an obviously territory ripe for more empirical effort. And, as observed above, there appears to be no obstacle to the analysis of the dependencies of total completion time. Finally,we suspect that there exist other arenas where strategic statistics like intercompletion times and total completion times can prove theoretically beneficial.